\documentclass[aps,prd,preprintnumbers,superscriptaddress,showpacs,floatfix]{revtex4}
\usepackage{epsfig}
\usepackage{bm}
\usepackage{amssymb}
\usepackage{amsmath}
\usepackage{color}

\begin{document}

\title{Solutions of Conformal Israel-Stewart Relativistic Viscous Fluid
Dynamics}
\author{Hugo Marrochio}
\author{Jorge Noronha}
\affiliation{Instituto de F\'{\i}sica, Universidade de S\~{a}o Paulo, C.P. 66318,
05315-970 S\~{a}o Paulo, SP, Brazil}
\author{Gabriel S.\ Denicol}
\affiliation{Department of Physics, McGill University, 3600 University Street, Montreal,
QC, H3A 2T8, Canada}
\author{Matthew Luzum}
\affiliation{Department of Physics, McGill University, 3600 University Street, Montreal,
QC, H3A 2T8, Canada}
\affiliation{Lawrence Berkeley National Laboratory, Nuclear Science Division, MS 70R0319,
Berkeley, CA 94720, USA}
\author{Sangyong Jeon}
\affiliation{Department of Physics, McGill University, 3600 University Street, Montreal,
QC, H3A 2T8, Canada}
\author{Charles Gale}
\affiliation{Department of Physics, McGill University, 3600 University Street, Montreal,
QC, H3A 2T8, Canada}

\begin{abstract}
We use symmetry arguments developed by Gubser to construct the first
radially-expanding explicit solutions of the Israel-Stewart formulation of
hydrodynamics. Along with a general semi-analytical solution, an exact
analytical solution is given which is valid in the cold plasma limit where
viscous effects from shear viscosity and the relaxation time coefficient are
important. The radially expanding solutions presented in this paper can be
used as nontrivial checks of numerical algorithms employed in hydrodynamic
simulations of the quark-gluon plasma formed in ultra-relativistic heavy ion
collisions. We show this explicitly by comparing such analytic and
semi-analytic solutions with the corresponding numerical solutions obtained
using the \textsc{music} viscous hydrodynamics simulation code.
\end{abstract}

\pacs{12.38.Mh, 47.75.+f, 47.10.ad, 11.25.Hf}
\maketitle
\date{\today }


\section{Introduction}

Our current understanding of the novel properties displayed by the
Quark-Gluon Plasma (QGP) formed in ultra-relativistic heavy ion collisions 
\cite{Gyulassy:2004zy} relies heavily on solving relativistic dissipative
fluid dynamics \cite{Heinz:2013th,gale}. The equations of relativistic fluid
dynamics form a set of complicated nonlinear partial differential equations,
describing the conservation of energy-momentum and a conserved charge (such
as net baryon number),%
\begin{equation*}
\partial _{\mu }T^{\mu \nu }=0,\text{ }\partial _{\mu }N^{\mu }=0\text{ }.
\end{equation*}%
In the presence of dissipation, the above equations are not closed and have
to be supplemented by nine additional equations of motion, i.e., the time
evolution equations for the the bulk viscous pressure, heat flow, and
shear-stress tensor.

The simplest formulation of relativistic dissipative fluid dynamics are the
Navier-Stokes equations. However, due to instabilities and acausal signal
propagation in these equations \cite{his}, they are not usually used in numerical
simulations. Currently, most fluid-dynamical simulations of the QGP employ
the relaxation-type equations derived by Israel and Stewart \cite%
{Israel:1979wp} to close the conservation laws. While some analytic
solutions of the non-relativistic Navier-Stokes equations are widely known 
\cite{landau}, very few analytical (or semi-analytical) solutions of
relativistic fluid dynamics \cite%
{Bjorken:1982qr,Csorgo:2003rt,Csorgo:2003ry,Csorgo:2006ax} have been
obtained and most simulations of heavy ion collisions solve dissipative
fluid dynamics numerically. Clearly, it would be useful to have solutions of
Israel-Stewart theory in any limit, especially in the cases relevant to
heavy ion collision applications.

In Refs.~\cite{gubser1,gubseryarom}, Gubser and Yarom derived $\mathrm{{SO(3)%
}\otimes {SU(1,1)}\otimes {Z}_{2}}$ invariant solutions of ideal
relativistic conformal fluid dynamics and relativistic Navier-Stokes theory.
In this paper we use the same symmetry arguments to derive solutions of
Israel-Stewart theory, which can be relevant to the description of the QGP.
Like the well known Bjorken solution \cite{Bjorken:1982qr}, the fluid
dynamic variables in the dissipative solutions we obtain are invariant under
Lorentz boosts in one direction, and are appropriate for comparison to data
from heavy-ion collisions near mid-rapidity, which are approximately
invariant under limited boosts in the beam direction. However, unlike the
Bjorken solution, they also have nontrivial radial expansion.

Thus, the solutions found in this paper provide the most rigorous tests to
date for the current numerical algorithms used to solve the viscous
relativistic fluid dynamic equations in heavy ion collisions. We show this
explicitly by comparing multi-dimensional numerical solutions obtained using 
\textsc{music}, a 3+1D viscous hydrodynamics simulation code \cite{MUSIC},
with the analytical and semi-analytical solutions of Israel-Stewart-like
theories undergoing Gubser flow. We remark that the version of \textsc{music}
employed in this work is an updated version currently being maintained at
McGill University.

This paper is organized as follows. In the next section we briefly introduce
the equations of relativistic dissipative fluid dynamics and describe the
solution for the flow velocity obtained by Gubser. In Sec.\ \ref{SecII} we
derive the main results of this paper and solve the equations of motion of
Israel-Stewart theory undergoing Gubser flow. We show in Sec.\ \ref{SecIII}
how these solutions can be used to test numerical simulations of
relativistic fluid dynamics. We conclude with a summary of our results.

\section{Hydrodynamics for heavy ion collisions and Gubser Flow}

\label{SecI}

In ultra-relativistic heavy ion applications, relativistic fluid dynamics is
more naturally described in hyperbolic coordinates $x^{\mu }=(\tau ,r,\phi
,\xi )$ where the line element is $ds^{2}=-d\tau ^{2}+dr^{\,2}+r^{2}d\phi
^{2}+\tau ^{2}d\xi ^{2}$, $r=\sqrt{x^{2}+y^{2}}$, and $\phi $ parametrize
the transverse plane perpendicular to the beam direction, while $\tau =\sqrt{%
t^{2}-z^{2}}$ and the rapidity $\xi =1/2\times \ln \left[ \left( t+z\right)
/\left( t-z\right) \right] $ are given in terms of usual coordinates $t$ and
the beam direction $z$.

The minimum set of relaxation-type equations for a viscous conformal fluid
is \cite{Baier:2007ix} 
\begin{eqnarray}
\frac{D_{\tau }T}{T}+\frac{\theta }{3}+\frac{\pi _{\mu \nu }\sigma ^{\mu \nu
}}{3sT} &=&0\,,  \label{energyeq} \\
\frac{\Delta _{\alpha }^{\mu }\nabla ^{\alpha }T}{T}+D_{\tau }u^{\mu }+\frac{%
\Delta _{\nu }^{\mu }\nabla _{\alpha }\pi ^{\alpha \nu }}{sT} &=&0\,,
\label{eulerrel} \\
\frac{\tau _{R}}{sT}\left( \Delta _{\alpha }^{\mu }\Delta _{\beta }^{\nu
}\,D_{\tau }\pi ^{\alpha \beta }+\frac{4}{3}\pi ^{\mu \nu }\theta \right) +%
\frac{\pi ^{\mu \nu }}{sT} &=&-\frac{2\eta }{s}\frac{\sigma ^{\mu \nu }}{T},
\label{pieq}
\end{eqnarray}%
where $\nabla _{\mu }$ is the space-time covariant derivative, $T$ is the
local temperature, $u^{\mu }$ is the 4-velocity of the fluid ($u_{\mu
}u^{\mu }=-1$), and $\pi ^{\mu \nu }$ is the shear-stress tensor. {We use
natural units $\hbar =c=k_{B}=1$. The metric tensor in flat spacetime is $%
g_{\mu \nu }=\mathrm{diag}\,(-,+,+,+)$}. We further introduced the entropy
density, $s\sim T^{3}$, the shear viscosity coefficient, $\eta $, the shear
relaxation time, $\tau _{R}$, the spatial projector, $\Delta _{\mu \nu
}\equiv g_{\mu \nu }+u_{\mu }u_{\nu }$, the comoving derivative, $D_{\tau
}\equiv u^{\lambda }\nabla _{\lambda }$, the expansion rate, $\theta \equiv
\nabla _{\alpha }u^{\alpha }$, and the shear tensor, $\sigma ^{\mu \nu
}\equiv \Delta ^{\mu \nu \alpha \beta }\nabla _{\alpha }u_{\beta }$, with $%
\Delta ^{\mu \nu \alpha \beta }\equiv \left( \Delta ^{\mu \alpha }\Delta
^{\nu \beta }+\Delta ^{\mu \beta }\Delta ^{\nu \alpha }\right) /2-\Delta
^{\mu \nu }\Delta ^{\alpha \beta }/3$ being the double, symmetric, traceless
projection operator. Even though other terms can be included in the
dynamical equation for the shear-stress tensor \cite%
{Baier:2007ix,Noronha:2011fi}, for simplicity in this paper we consider only
the terms present in (\ref{pieq}).

Equation (\ref{pieq}) contains two transport coefficients, $\eta $ and $\tau
_{R}$. In a conformal fluid, the shear viscosity coefficient is always
proportional to the entropy density, $\eta \sim s$, while the shear
relaxation time must be proportional to the inverse of the temperature, $%
\tau _{R}\sim 1/T$. Without loss of generality, the relaxation time is
parametrized as, $\tau _{R}=c\,\eta /(Ts)$, where $c$ is a constant.

We shall consider here the case in which the dynamics is boost invariant and
the flow is radially symmetric, i.e., $T=T(\tau ,r)$ and $\pi ^{\mu \nu
}=\pi ^{\mu \nu }(\tau ,r)$. These conditions are approximately met near
mid-rapidity in ultra--central collisions at the LHC, recently measured by
the ATLAS and CMS Collaborations \cite{ATLAS:2012at, CMS:2012xxa}. In fact,
we will follow \cite{gubser1} and assume that the conformal fluid flow is
actually invariant under $\mathrm{{SO(3)}\otimes {SO(1,1)}\otimes {Z}_{2}}$.
The $\mathrm{SO(3)}$ piece is a subgroup of the $\mathrm{SO(4,2)}$ conformal
group which describes the symmetry of the solution under rotations around
the beam axis and two operations constructed using special conformal
transformations that replace translation invariance in the transverse plane.
For more details regarding the generators of the $\mathrm{SO(3)}$ symmetry
group of this solution, see Ref.\ \cite{gubser1}. The $\mathrm{{Z}_{2}}$
piece stands for invariance under $\xi \rightarrow -\xi $, while $\mathrm{%
SO(1,1)}$ denotes invariance under boosts along the beam axis. In this case,
the dynamical variables depend on $\tau $ and $r$ through the dimensionless
combination \cite{gubser1,gubseryarom} 
\begin{equation}
g(\tilde{\tau},\tilde{r})=\frac{1-\tilde{\tau}^{2}+\tilde{r}^{2}}{2\tilde{%
\tau}},
\end{equation}%
where $\tilde{\tau}\equiv q\tau $ and $\tilde{r}=qr$, with $q$ being an
arbitrary energy scale. The flow is completely determined by symmetry
constraints to be \cite{gubser1,gubseryarom}%
\begin{eqnarray}
u_{\tau } &=&-\cosh \left[ \tanh ^{-1}\left( \frac{2\tilde{\tau}\tilde{r}}{1+%
\tilde{\tau}^{2}+\tilde{r}^{2}}\right) \right] \,,  \notag \\
u_{r} &=&\sinh \left[ \tanh ^{-1}\left( \frac{2\tilde{\tau}\tilde{r}}{1+%
\tilde{\tau}^{2}+\tilde{r}^{2}}\right) \right] \,,  \notag \\
u_{\phi } &=&u_{\xi }=0\,.  \label{GubserFlow}
\end{eqnarray}%
In the following, this solution will be referred to as Gubser flow. Since
the flow is known, the relativistic Euler equation (\ref{eulerrel}) is
automatically satisfied and, thus, only the equations for the temperature (%
\ref{energyeq}) and the shear-stress tensor (\ref{pieq}) need to be solved.

The dynamical fields can be written in terms of the dimensionless variables $%
\tilde{T}(g)\equiv T(\tau ,r)/q$ and $\tilde{\pi}^{\mu \nu }(g)\equiv \pi
^{\mu \nu }(\tau ,r)/q^{4}$, which converts the partial differential
equations into simple ordinary differential equations for these variables.
The other variables become $\tilde{\theta}(g)=\theta /q$, $\tilde{\sigma}%
_{\mu \nu }(g)=\sigma _{\mu \nu }/q\,$, $\tilde{\tau}_{R}=c(\eta /s)/\tilde{T%
}$, $s=\alpha q^{3}\,\tilde{T}^{3}$ and the nontrivial equations of motion
considerably simplify to%
\begin{eqnarray}
\frac{1}{\tilde{T}(g)}D_{\tilde{\tau}}\tilde{T}(g)+\frac{1}{3}\tilde{\theta}%
(g)+\frac{\tilde{\pi}_{\mu \nu }(g)\tilde{\sigma}^{\mu \nu }(g)}{3\alpha 
\tilde{T}(g)^{4}} &=&0\,,  \label{energyeqnew} \\
\tilde{\tau}_{R}\left[ \Delta _{\alpha }^{\mu }\Delta _{\beta }^{\nu }\,D_{%
\tilde{\tau}}\tilde{\pi}^{\alpha \beta }(g)+\frac{4}{3}\tilde{\pi}^{\mu \nu
}(g)\tilde{\theta}(g)\right] +\tilde{\pi}^{\mu \nu }(g) &=&-2\,\eta \tilde{%
\sigma}^{\mu \nu }(g)\,.  \label{pieqnew}
\end{eqnarray}

We now follow \cite{gubseryarom} and perform a Weyl rescaling so then the
expanding fluid in hyperbolic coordinates becomes a static fluid in some
other, conveniently chosen, expanding coordinate system. The dimensionless
line element in hyperbolic coordinates can be written as $d\tilde{s}^{2}=-d%
\tilde{\tau}^{2}+d\tilde{r}^{\,2}+\tilde{r}^{2}d\phi ^{2}+\tilde{\tau}%
^{2}d\xi ^{2}$. By performing a Weyl rescaling of this metric we obtain $d%
\tilde{s}^{2}\rightarrow d\hat{s}^{2}\equiv d\tilde{s}^{2}/\tau^2$, which is the metric in $\mathrm{dS%
}_{3}\otimes \mathbf{R}$, where $\mathrm{dS}_{3}$ corresponds to the
3-dimensional de Sitter space. A convenient coordinate transformation
introduced in \cite{gubseryarom} takes $d\hat{s}^{2}$ to $d\hat{s}%
^{2}=-d\rho ^{2}+\cosh ^{2}\rho \,d\theta ^{2}+\cosh ^{2}\rho \sin
^{2}\theta \,d\phi ^{2}+d\xi ^{2}$, where 
\begin{equation}
\sinh \rho =-\frac{1-\tilde{\tau}^{2}+\tilde{r}^{2}}{2\tilde{\tau}}\,,\qquad
\tan \theta =\frac{2\tilde{r}}{1+\tilde{\tau}^{2}-\tilde{r}^{2}}\,.
\label{definerhoeq}
\end{equation}%
In the following, we denote all fluid-dynamical variables in this coordinate
system with a hat. Such generalized de Sitter coordinate is extremely
convenient since it leads to a static velocity profile, i.e., $\hat{u}_{\mu
}=(-1,0,0,0)$, and considerably simplifies the calculations. Since $g=-\sinh
\rho $, the fields are only functions of $\rho $, i.e, $\hat{T}=\hat{T}(\rho
)$ and $\hat{\pi}^{\mu \nu }=\hat{\pi}^{\mu \nu }(\rho )$. Because of the
metric rescaling and the coordinate transformation $\tilde{x}^{\mu }=(\tilde{%
\tau},\tilde{r},\phi ,\xi )\rightarrow \hat{x}^{\mu }(\rho ,\theta ,\phi
,\xi )$, the dimensionless dynamical variables in $\mathrm{dS}_{3}\otimes 
\mathbf{R}$ are related to those in hyperbolic coordinates as follows 
\begin{eqnarray}
u_{\mu }(\tilde{\tau},\tilde{r}) &=&\tilde{\tau}\frac{\partial \hat{x}^{\nu }%
}{\partial \tilde{x}^{\mu }}\hat{u}_{\nu }\,,  \label{transformu} \\
\tilde{T}(\tilde{\tau},\tilde{r}) &=&\frac{\hat{T}}{\tilde{\tau}}\,,
\label{transformf} \\
\tilde{\pi}_{\mu \nu }(\tilde{\tau},\tilde{r}) &=&\frac{1}{\tilde{\tau}^{2}}%
\frac{\partial \hat{x}^{\alpha }}{\partial \tilde{x}^{\mu }}\frac{\partial 
\hat{x}^{\beta }}{\partial \tilde{x}^{\nu }}\hat{\pi}_{\alpha \beta }\,.
\label{transformh}
\end{eqnarray}%
The factors of $\tilde{\tau}$ in the transformation rules above come from
the known properties of these fields under Weyl transformations \cite%
{gubseryarom}. For instance, since $\pi _{\mu \nu }\rightarrow \Omega
^{2}\pi _{\mu \nu }$ under Weyl rescaling $g_{\mu \nu }\rightarrow \Omega
^{-2}\,g_{\mu \nu }$ with $\Omega =\tilde{\tau}$, there is a factor of $1/%
\tilde{\tau}^{2}$ in (\ref{transformh}). Given the dictionary between the
fields in the different spaces shown above, one can solve the equations (\ref%
{energyeqnew}) and (\ref{pieqnew}) in $\mathrm{dS}_{3}\otimes \mathbf{R}$
where the fluid is static and the fields are homogeneous (i.e., they only
depend on the de Sitter time coordinate $\rho $) and plug in the solutions
to find the fields in the standard flat space-time. This is the general
strategy that we shall follow below to find solutions for the viscous
relativistic fluid defined above.

\section{Israel-Stewart theory}

\label{SecII}

In this section we derive for the first time the solutions of Israel-Stewart
theory in the Gubser flow regime. These solutions shall be later compared to
numerical simulations of fluid dynamics. It is straightforward to work out
the equations of motion for $\hat{T}(\rho )$ and $\hat{\pi}_{\nu }^{\mu
}(\rho )$ in the generalized de Sitter coordinates. First, note that
orthogonality to the flow gives $\hat{\pi}_{\rho }^{\mu }=0$ (where $\mu
=\rho ,\theta ,\phi ,\xi $) while the tracelessness condition imposes $\hat{%
\pi}_{\xi }^{\xi }=-\hat{\pi}_{\theta }^{\theta }-\hat{\pi}_{\phi }^{\phi }$%
. Since the only nonzero components of the shear tensor are $\hat{\sigma}%
_{\xi }^{\xi }=-2\tanh \rho /3$, $\hat{\sigma}_{\theta }^{\theta }=\hat{%
\sigma}_{\phi }^{\phi }=\tanh \rho /3$, each one of the off diagonal terms
of the $\hat{\pi}_{\nu }^{\mu }$ tensor follows an independent, $1$--st
order linear homogeneous equation and we set their initial conditions to
zero (thus, they do not contribute to the dynamics). One can show that $\hat{%
\pi}_{\theta }^{\theta }$ and $\hat{\pi}_{\phi }^{\phi }$ obey the same
differential equations and, since we impose the same initial conditions for
these fields, $\hat{\pi}_{\xi }^{\xi }=-2\hat{\pi}_{\theta }^{\theta }=-2%
\hat{\pi}_{\phi }^{\phi }$. We then find that the only nontrivial nonlinear
equations of motion are

\begin{eqnarray}
\frac{1}{\hat{T}}\frac{d\hat{T}}{d\rho }+\frac{2}{3}\tanh \rho &=&\frac{1}{3}%
\bar{\pi}_{\xi }^{\xi }(\rho )\,\tanh \rho \,,  \label{finalequations1} \\
\frac{c}{\hat{T}}\frac{\eta }{s}\left[ \frac{d\bar{\pi}_{\xi }^{\xi }}{d\rho 
}+\frac{4}{3}\left( \bar{\pi}_{\xi }^{\xi }\right) ^{2}\tanh \rho \right] +%
\bar{\pi}_{\xi }^{\xi } &=&\frac{4}{3}\frac{\eta }{s\hat{T}}\tanh \rho \,,
\label{finalequations2}
\end{eqnarray}
where $\bar{\pi}_{\xi }^{\xi }\equiv \hat{\pi}_{\xi }^{\xi }/(\hat{T}\hat{s}%
) $. This variable is convenient since it is invariant under Weyl
transformations. In order to derive the equations above we used that $\hat{%
\theta}=2\tanh \rho $.

Note that for any nonzero $\tau $, the value of $\rho $ decreases with $r$,
while for a fixed $r$ the value of $\rho $ increases with $\tau $. Thus,
when $\rho \ll 0$ one probes regions in which $r\gg 1$, and when $\rho \gg 1$
one has $\tau \gg 1$. In this sense, we expect that physically meaningful
solutions behave as $\lim_{\rho \rightarrow \pm \infty }\hat{T}(\rho )=0$,
i.e., at an infinite radius or time the temperature should go to zero. On
the other hand, given the definition of $\bar{\pi}_{\xi }^{\xi }$, it is
consistent to have $\lim_{\rho \rightarrow \pm \infty }\bar{\pi}_{\xi }^{\xi
}(\rho )$ finite and nonzero ($\bar{\pi}_{\xi }^{\xi }$ is a ratio between
two quantities that should vanish when $\rho \rightarrow \pm \infty $).

When $\bar{\pi}_{\xi }^{\xi }=0$, we have only a single equation left over
for the temperature and the analytical solution is the one found in \cite%
{gubser1,gubseryarom} 
\begin{equation}
\hat{T}_{\mathrm{ideal}}(\rho )=\frac{\hat{T}_{0}}{\cosh ^{2/3}\rho }\,,
\label{solutionTideal}
\end{equation}%
where $\hat{T}_{0}\equiv \hat{T}_{\mathrm{ideal}}(0)$ is a positive constant
(so then $\hat{T}_{\mathrm{ideal}}$ is positive-definite). Using the
dictionary in (\ref{transformf}), we see that the temperature in the
original hyperbolic coordinates is given by 
\begin{equation}
T_{\mathrm{ideal}}(\tau ,r)=\frac{\hat{T}_{0}(2q\tau )^{2/3}}{\tau \left[
1+2q^{2}(\tau ^{2}+r^{2})+q^{4}(\tau ^{2}-r^{2})^{2}\right] ^{1/3}},
\end{equation}%
and, at the time $\tau _{0}=1/q$, one finds $T_{\mathrm{ideal}}(\tau _{0},0)=%
\hat{T}_{0}\,q$.

The relativistic Navier-Stokes approximation to our set of equations
consists in setting $c=0$ (i.e., the relaxation time coefficient is set to
zero) while keeping $\eta /s$ nonzero in (\ref{finalequations2}). In this
case, $\bar{\pi}_{\xi }^{\xi }(\rho )=4/(3\hat{T})\times (\eta /s)\tanh \rho 
$ and the equation for $\hat{T}$ becomes 
\begin{equation*}
\frac{d}{d\rho }\hat{T}+\frac{2}{3}\hat{T}\tanh \rho =\frac{4}{9}\frac{\eta 
}{s}\left( \tanh \rho \right) ^{2}.\,
\end{equation*}%
The analytical solution, previously found in \cite{gubser1,gubseryarom}, is 
\begin{equation}
\hat{T}_{\mathrm{NS}}(\rho )=\frac{\hat{T}_{0}}{\cosh ^{2/3}\rho }+\frac{4}{%
27}\frac{\eta }{s}\frac{\sinh ^{3}\rho }{\cosh ^{2/3}\rho }%
\,\,_{2}F_{1}\left( \frac{3}{2},\frac{7}{6};\frac{5}{2};-\sinh ^{2}\rho
\right) ,  \label{NavierStokesTsolution}
\end{equation}%
where $_{2}F_{1}$ is a hypergeometric function. From the equation of motion,
the condition $\lim_{\rho \rightarrow \pm \infty }\hat{T}_{\mathrm{NS}%
}^{\prime }(\rho )=0$ shows that $\lim_{\rho \rightarrow \pm \infty }\hat{T}%
_{\mathrm{NS}}(\rho )=\pm 2\eta /3s$ \cite{gubser1,gubseryarom}. In this
case, once $\eta /s\neq 0$, for any given $\tau $ there is always a value of 
$r$ beyond which the temperature switches sign and becomes negative (which
is very different than the ideal case in which $\lim_{\rho \rightarrow \pm
\infty }\hat{T}_{\mathrm{ideal}}=0$). This effect may be connected with the
well-known causality issue (see, for instance, \cite%
{Denicol:2008ha,Pu:2009fj}) of the relativistic Navier-Stokes equations. We
shall see below that once the relaxation time coefficient is taken into
account one can find a solution where $\hat{T}$ is positive-definite and $%
\lim_{\rho \rightarrow -\infty }\hat{T}(\rho )=0$.

Obtaining solutions of Israel-Stewart theory is more evolved, since the
relaxation time in Eq.\ (\ref{finalequations2}) cannot be set to zero, i.e., 
$\tau _{R}\neq 0$. In this case $\bar{\pi}_{\xi }^{\xi }$ obeys a
differential equation (which requires an independent initial condition) and
the set of equations becomes nonlinear. At the very least, it is possible to
find one qualitative difference between the asymptotic solutions ($%
\lim_{\rho \rightarrow \pm \infty }$) of Navier-Stokes and Israel-Stewart
theories. If one imposes that $\lim_{\rho \rightarrow \pm \infty }\hat{T}%
(\rho )=0$ and, simultaneously, $\lim_{\rho \rightarrow \pm \infty }d\bar{\pi%
}_{\xi }^{\xi }(\rho )/d\rho =0$, one can find the asymptotic solution for $%
\bar{\pi}_{\xi }^{\xi }(\rho )$, $\lim_{\rho \rightarrow \pm \infty }|\bar{%
\pi}_{\xi }^{\xi }(\rho )|=\sqrt{1/c}$. Therefore, in contrast to
Navier-Stokes theory solutions in which $\lim_{\rho \rightarrow \pm \infty }%
\hat{T}(\rho )=0$ are possible in Israel-Stewart theory and do happen in
practice as long as $\tau_R$ is nonzero.

There is a limit in which one can find analytical solutions for $\hat{T}$
and $\bar{\pi}_{\xi }^{\xi }$. This becomes possible when the fluid is very
viscous or when the temperature is very small, i.e., when $\eta /(s\hat{T}%
)\gg 1$. In this case, called here the \textit{cold plasma limit}, the term $%
\bar{\pi}_{\xi }^{\xi }$ becomes negligible in comparison to all the other
terms in Eq.\ (\ref{finalequations2}), which are all linear in $\eta /(s\hat{%
T})$. In this limit, one can directly solve the equation for $\bar{\pi}_{\xi
}^{\xi }$ to find 
\begin{equation}
\bar{\pi}_{\xi }^{\xi }(\rho )=\sqrt{\frac{1}{c}}\tanh \left[ \sqrt{\frac{1}{%
c}}\left( \frac{4}{3}\,\ln \,\cosh \rho -\bar{\pi}_{0}c\right) \right] ,
\label{analh}
\end{equation}%
where $\bar{\pi}_{0}$ is a constant and, substituting this into Eq.\ (\ref%
{finalequations1}), we obtain 
\begin{equation}
\hat{T}(\rho )=\hat{T}_{1}\frac{\exp \left( c\bar{\pi}_{0}/2\right) }{(\cosh
\rho )^{2/3}}\cosh ^{1/4}\left[ \sqrt{\frac{1}{c}}\left( \frac{4}{3}\,\ln
\,\cosh \rho -\bar{\pi}_{0}c\right) \right] \,.  \label{analf}
\end{equation}%
where $\hat{T}_{1}$ is a constant. These analytical solutions are even in $%
\rho $, $\hat{T}$ is positive-definite, and $\lim_{\rho \rightarrow \pm
\infty }\hat{T}(\rho )=0$ if $4c>1$. Moreover, note that as long as $c>1$, $%
\bar{\pi}_{\xi }^{\xi }$ is always smaller than 1 for any value of $\rho $,
i.e, the dissipative correction to the energy-momentum tensor is always
smaller than the ideal fluid contribution. In the next section, the
analytical solutions in Eqs.\ (\ref{analh}) and (\ref{analf}) will be
compared to numerical solutions of fluid dynamics obtained with \textsc{music%
}.

We show in Fig.\ 1 a comparison between $\hat{T}$ and $\bar{\pi}_{\xi }^{\xi
}$ computed for an ideal fluid, Navier-Stokes theory, and
Israel-Stewart theory for $\eta /s=0.2$, which is a value in the
ballpark of that normally used in hydrodynamic simulations of the QGP in
heavy ion collisions~\cite{Luzum:2012wu}, and $c=5$, which is the typical
value obtained from approximations of the Boltzmann equation \cite%
{Denicol:2010xn,Denicol:2011fa,Denicol:2012cn}. We have chosen the initial
conditions for the equations such that $\hat{T}(0)=1.2$, for all the cases,
and, for the Israel-Stewart case, $\bar{\pi}_{\xi }^{\xi }(0)=0$. We solve
Eqs.\ (\ref{finalequations1}) and (\ref{finalequations2}) numerically using 
\textsc{mathematica}'s NDSolve subroutine. The Israel-Stewart theory results
are shown in solid black, the Navier-Stokes results in dashed blue, and the
ideal fluid result in the dashed-dotted red curve. One can see that the
Israel-Stewart solution for $\hat{T}$ is positive-definite and $\lim_{\rho
\rightarrow \pm \infty }\hat{T}(\rho )=0$. Moreover, viscous effects break
the parity of the solutions with respect to $\rho \rightarrow -\rho $. Note
that, as mentioned before, $\bar{\pi}_{\xi }^{\xi }$ goes to $\sqrt{1/c}$
when $\rho \rightarrow \pm \infty $ in Israel-Stewart theory while for the
Navier-Stokes solution this quantity diverges at $\rho \approx -4.19$, which
is the value of $\rho $ at which $\hat{T}_{NS}=0$. We also checked that the
analytical limit in Eqs.\ (\ref{analh}) and (\ref{analf}) matches
the numerical solution for $\eta /s=1/(4\pi )$ \cite{Kovtun:2004de} and $c=5$
when $\hat{T}(0)\leq 0.001$, i.e., when the temperature is extremely small. 
\begin{figure}[tbp]
\begin{minipage}{.4\linewidth}
\hspace{-1.5cm}
\includegraphics[width=8.2cm]{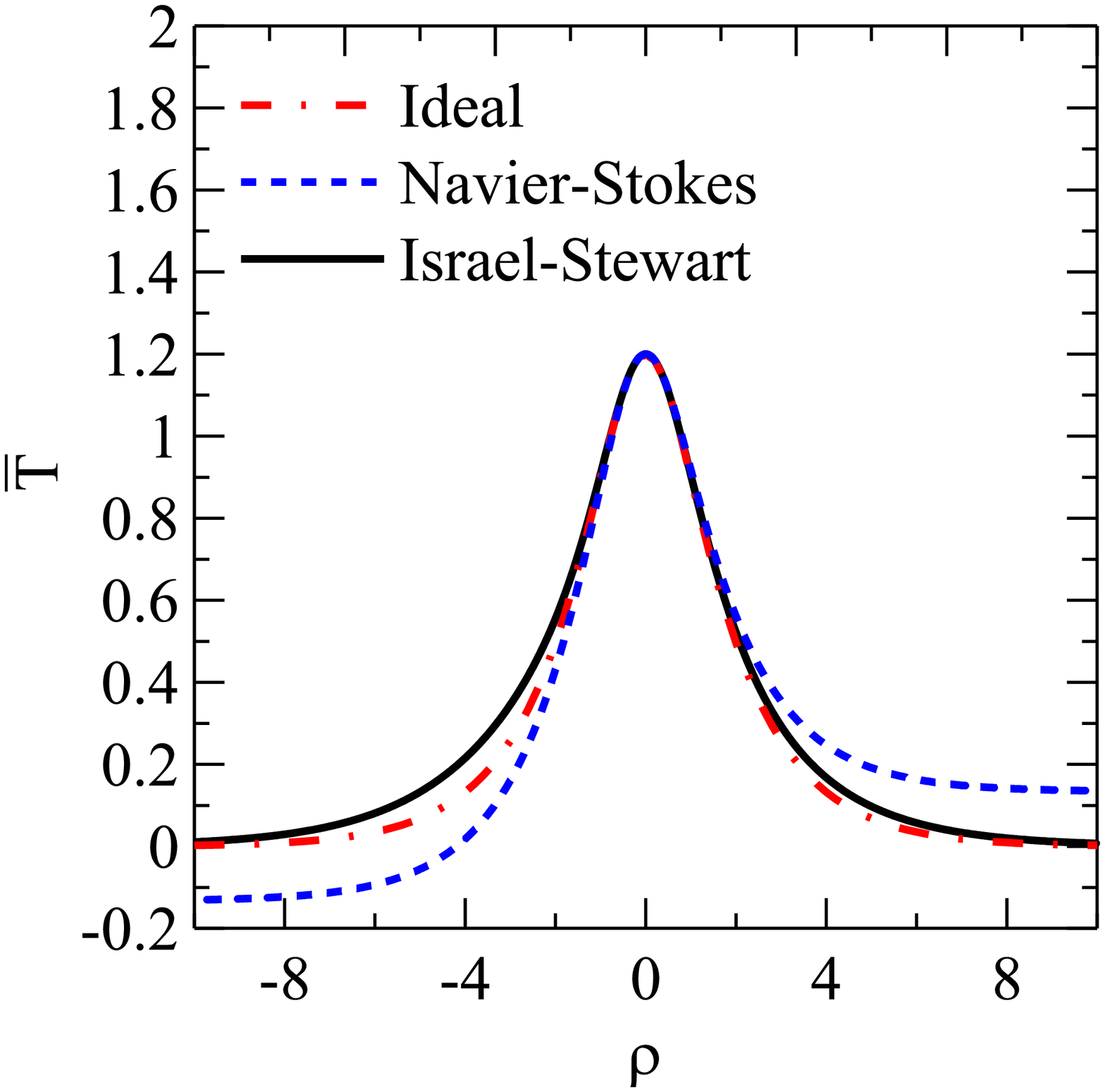} 
\end{minipage}  
\begin{minipage}{.4\linewidth}
\hspace{-1.5cm}
\includegraphics[width=8.2cm]{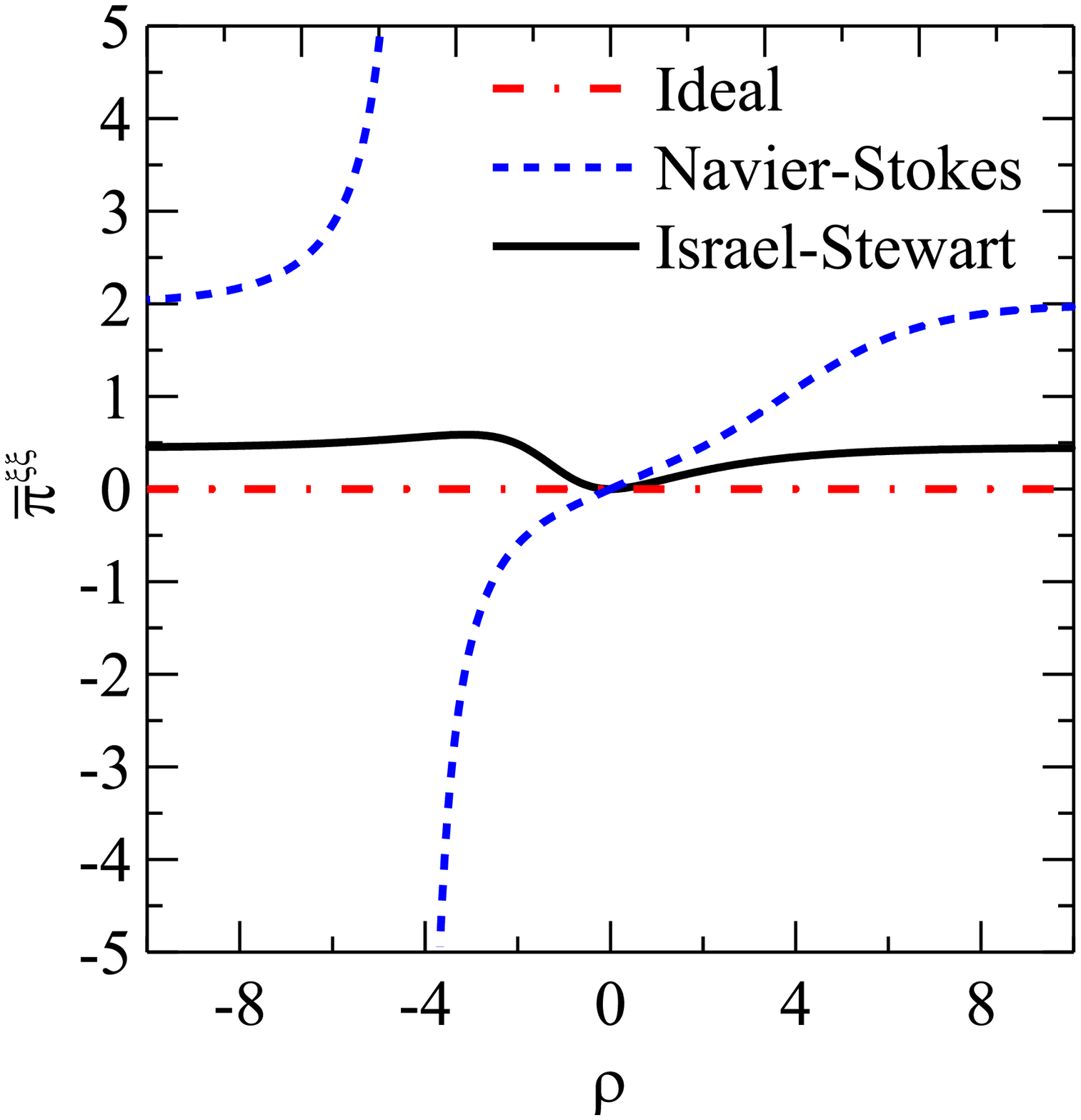} 
\end{minipage} 
\caption{Comparison between the solutions for $\hat{T}$ (left panel) and $\bar{\protect\pi%
}_{\protect\xi }^{\protect\xi }$ (right panel) for $\protect\eta /s=0.2$, $c=5$, and $\hat{%
T}(0)=1.2$ found using different versions of the relativistic fluid
equations. The solid black lines denote solutions of Israel-Stewart theory, results from relativistic Navier-Stokes theory are in
dashed blue, while the dashed-dotted red curves correspond to the ideal
fluid case.}
\label{fig1}
\end{figure}

\begin{figure}[tbp]
\begin{minipage}{.4\linewidth}
\hspace{-1.5cm}
\includegraphics[width=8.2cm]{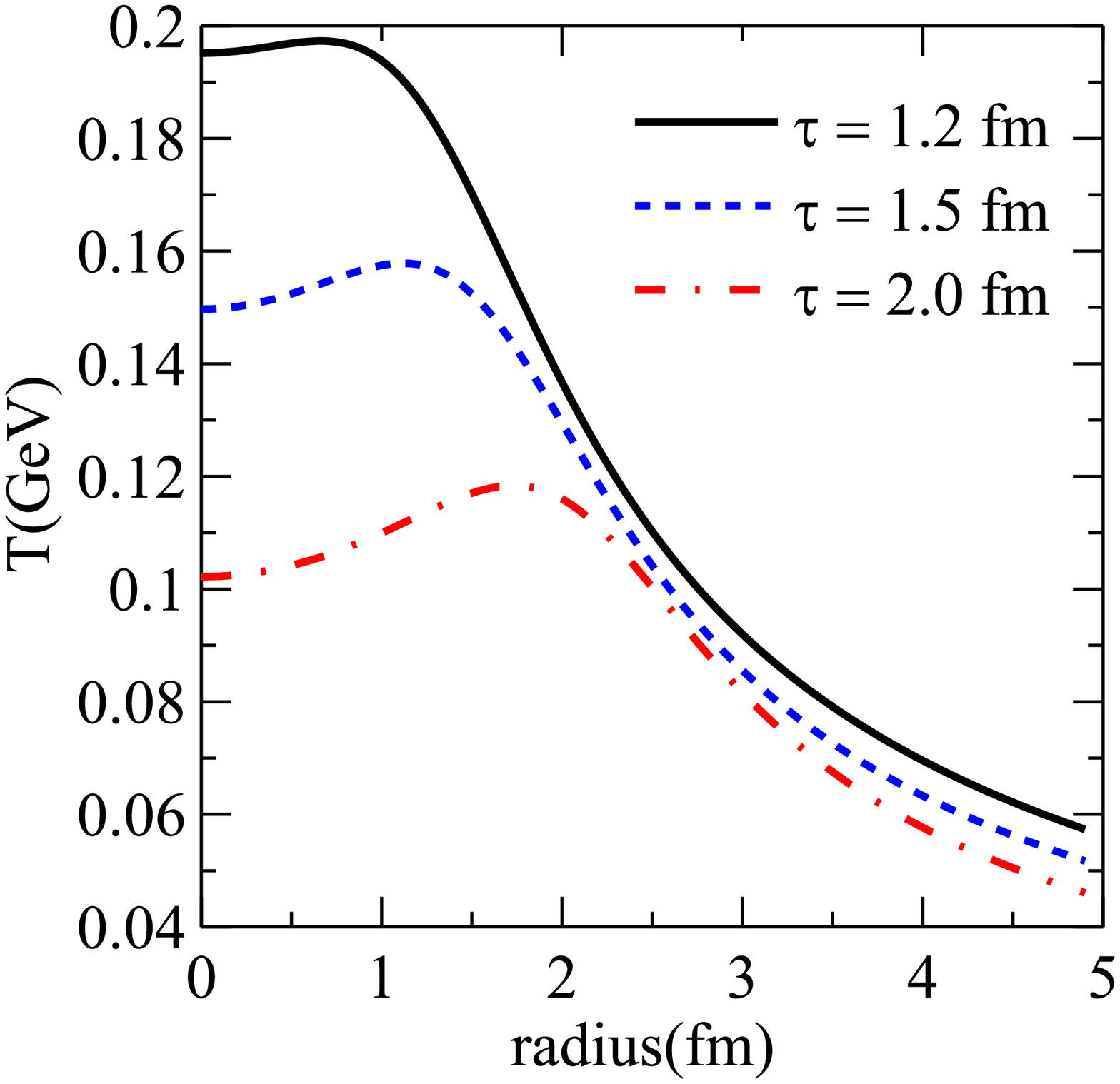} 
\end{minipage}  
\begin{minipage}{.4\linewidth}
\hspace{-1.5cm}
\includegraphics[width=8.2cm]{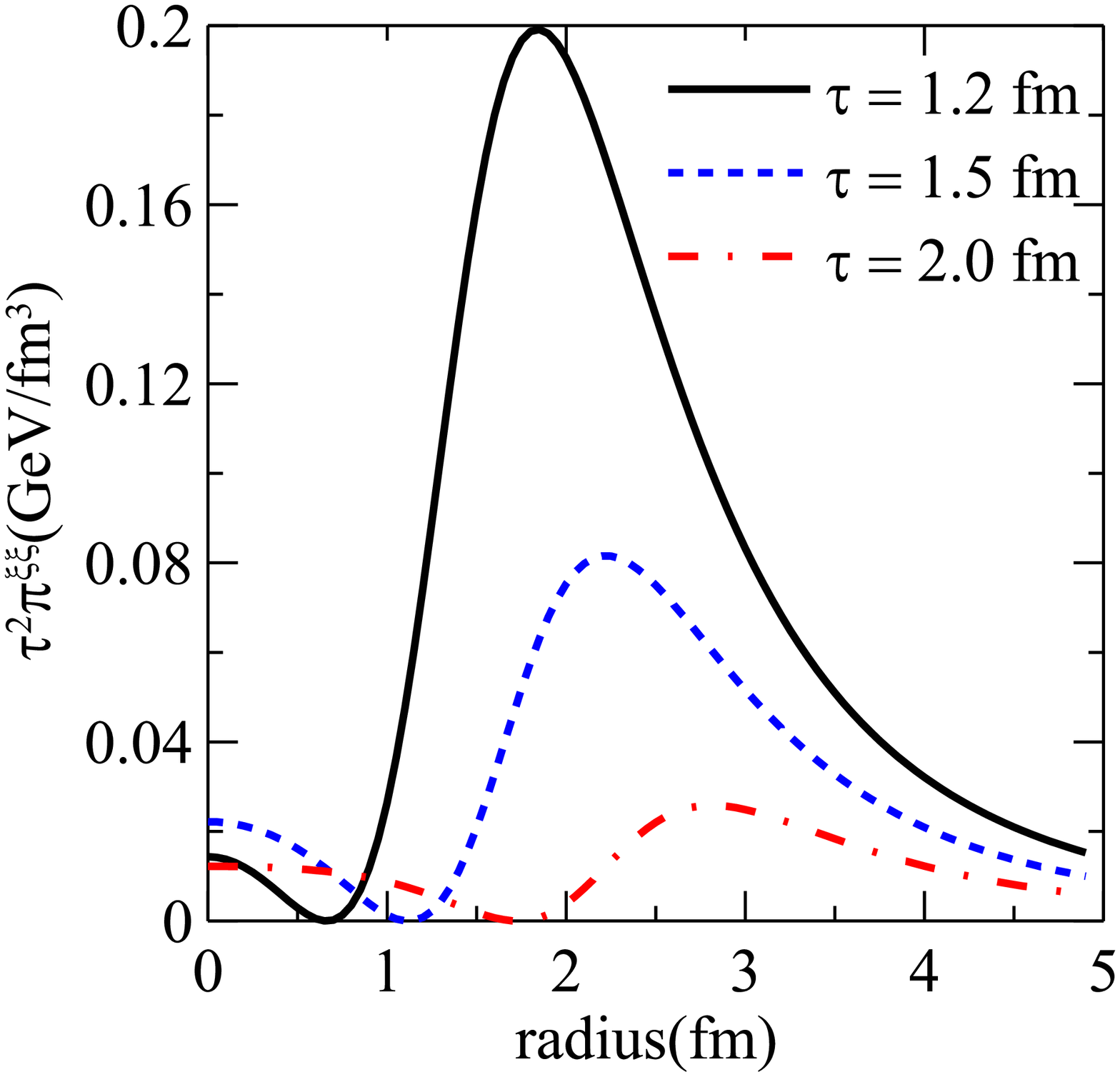} 
\end{minipage} 
\caption{Temperature and $\protect\tau ^{2}\protect\pi ^{\protect\xi \protect%
\xi }$ profiles in Israel-Stewart theory for $\protect\tau =1.2$ fm (solid
black curves), $\protect\tau =1.5$ fm (dashed blue curves), and $\protect%
\tau =2$ fm (dashed-dotted red curves) with $q=1$ fm$^{-1}$, $\protect\eta %
/s=0.2$, $c=5$, and $\hat{T}(0)=1.2$.}
\label{fig2}
\end{figure}

In order to study the space-time dependence of the Israel-Stewart solutions
we define $q=1$ fm$^{-1}$ so that $\rho =0$ corresponds to $\tau =1$ fm and $%
r=0$. Therefore, in standard hyperbolic coordinates, $T\left( r={0,\tau }%
_{0}=1\text{ fm}\right) =1.2$ fm$^{-1}$ and $\bar{\pi}_{\xi }^{\xi }\left( r=%
{0,\tau }_{0}=1\text{ fm}\right) =0$. In Fig.\ \ref{fig2} we show a
comparison between the temperature profiles for Israel-Stewart theory at the
times $\tau =1.2$, $1.5$, $2$ fm, with $\eta /s=0.2$, $c=5$. Also, in the
same figure we show $\tau ^{2}\pi ^{\xi \xi }$ as a function of the radius
for the same times. The other components of the shear-stress tensor can be
obtained using the dictionary in Eq.\ (\ref{transformh}). 

\section{Testing Fluid dynamics}

\label{SecIII}

While there are analytical and semi-analytical solutions of relativistic
ideal fluid dynamics \cite%
{Bjorken:1982qr,Csorgo:2003rt,Csorgo:2003ry,Csorgo:2006ax}, the same is not
the case for Israel-Stewart theory. This makes testing numerical algorithms
that solve the equations of relativistic fluid dynamics rather problematic.
Procedures such as to fix the numerical viscosity, choose the appropriate
parameters for the flux limiters, among others, which strongly rely on trial
and error based tests, become then highly nontrivial. Furthermore, most
algorithms used to numerically solve the equations of Israel-Stewart theory
were not developed for this purpose: they were developed to solve
conservation laws or even Navier-Stokes theory, usually in the
non-relativistic limit. In practice, most simulation codes used in heavy ion
collisions have to adapt such algorithms to also solve Israel-Stewart
theory. In this sense, the set of parameters that were found optimal to
solve certain problems in the non-relativistic regime, such as the Riemann
problem \cite{Rischke:1998fq}, might not be optimal to solve Israel-Stewart
theory in the conditions produced in heavy ion collisions.

In this section, we compare numerical solutions of dissipative fluid
dynamics obtained via the Kurganov-Tadmor (KT) algorithm \cite{KTscheme}
using \textsc{music} \cite{MUSIC}, with semi-analytical solutions of
(conformal) Israel-Stewart theory in the Gubser flow scenario. We show how
this can be used to probe not only the quality and accuracy of the dynamical
simulation but also to find the optimal value for some of the (numerical)
parameters that exist in the algorithm.

In the standard version of \textsc{music}, the evolution equations that are
solved are already those listed in Eqs.\ (\ref{energyeq}), (\ref{eulerrel}),
and (\ref{pieq}). Therefore, the solutions calculated with \textsc{music}
can already be compared with those of Gubser flow obtained in the previous
section. For a meaningful comparison, one must initialize the numerical
simulation with an initial condition constructed from the solutions of Eqs. (%
\ref{finalequations1}) and (\ref{finalequations2}) for a given initial time.
In this work, we fix the initial time to be $\tau _{0}=1$ fm. The
temperature at $\rho =0$, which determines the temperature at $\tau =\tau
_{0}$ and $r=0$, is fixed to be $T=T_{0}=1.2$ fm$^{-1}$. The shear-stress
tensor at $\rho =0 $ is initialized to be $\pi ^{\mu \nu }=0$. The viscosity
in \textsc{music} is set to $\eta /s=0.2$ while the relaxation time is fixed
to $\tau _{R }=5\eta /(\varepsilon +P)$, i.e., $c=5$. This parametrization
for the relaxation time guarantees that the fluid dynamical evolution is
causal \cite{Pu:2009fj}. The time step and grid spacing used in the
numerical simulation are $\delta \tau =0.005$ fm and $\delta x=\delta y=0.05$
fm, respectively ($\delta \tau$, $\delta x$, and $\delta y$ are small enough
to achieve the continuum limit). We remark that in Gubser flow the values of
the transport coefficients actually affect the initial condition of the
fluid, since in this scheme the initial condition in hyperbolic coordinates
must also be constructed by actually solving the fluid-dynamical equations
in the generalized de Sitter space.

Note that \textsc{music} was originally designed to solve Israel-Stewart
theory in 3+1--dimensions, while the Gubser flow solution assumes boost
invariance. In a numerical simulation in 3+1--dimensions, boost invariance
can be trivially obtained by providing an initial condition that is also
boost invariant. In this situation, the solutions of fluid dynamics should
maintain exact boost invariance, remaining trivial in the longitudinal
direction. We checked that this does occur in the solutions obtained with 
\textsc{music}: the temperature and $\pi ^{\mu \nu }$ profiles remain
(exactly) constant in the $\xi $--direction (e.g., $\pi ^{\xi x}$, $\pi
^{x\xi }$, $\pi ^{\xi y}$, $\pi ^{y\xi }$ are exactly zero) while the
longitudinal component of the velocity field is exactly zero. This is only
not the case at the boundary of the grid where boost invariance is not
exactly maintained due to finite size effects.

\subsection{Comparison to semi-analytical solution}

In the following we compare the numerical solutions of \textsc{music} with
the semi-analytical solutions of Israel-Stewart theory. Figures\ \ref%
{MUSIC_T_vel} and \ref{MUSIC_W} show the spatial profiles of temperature, $T$%
, velocity, $u^{x}$, and the $\xi \xi $, $yy$, and $xy$ components of the
shear-stress tensor, $\pi ^{\xi \xi }$, $\pi ^{yy} $, and $\pi ^{xy}$,
respectively. Without loss of generality, $T$, $u^{x}$, $\pi ^{\xi \xi }$, $%
\pi ^{yy}$ are shown as a function of $x$ in the $y=0$ axis, while the $\pi
^{xy}$ profile is shown as a function of $x$ in the $x=y $ direction. The
component $\pi ^{xy}$ vanishes on the $x$,$y$--axis, which we verified also
happens in \textsc{music}. Note that all the other components of $\pi ^{\mu
\nu }$ can be obtained from the $3$ components displayed, i.e., $\pi ^{\xi
\xi }$, $\pi ^{yy}$, and $\pi ^{xy}$.

\begin{figure}[tbp]
\begin{minipage}{.4\linewidth}
\hspace{-1.5cm}
\includegraphics[width=8.2cm]{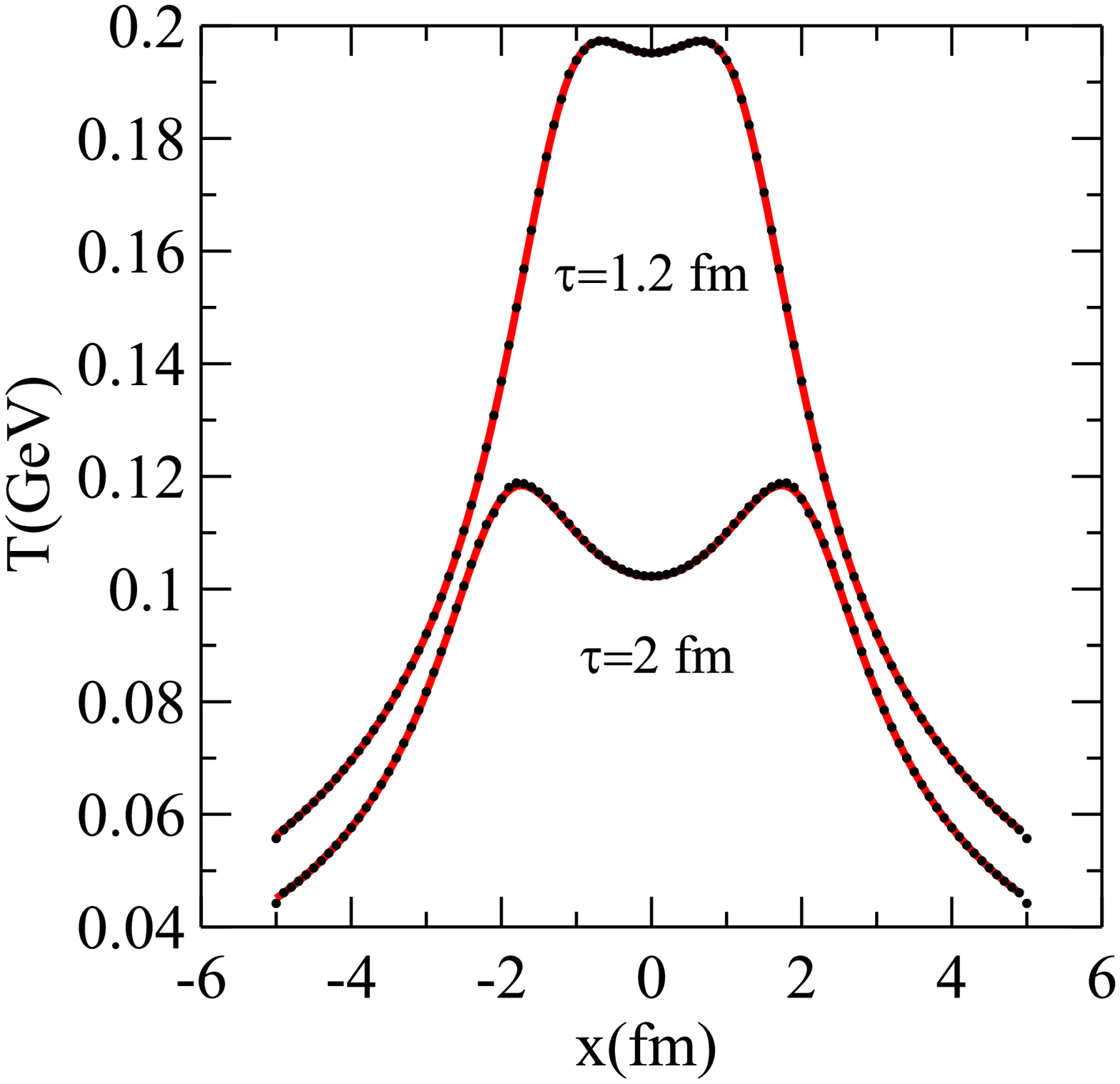} 
\end{minipage}  
\begin{minipage}{.4\linewidth}
\hspace{-1.5cm}
\includegraphics[width=8.2cm]{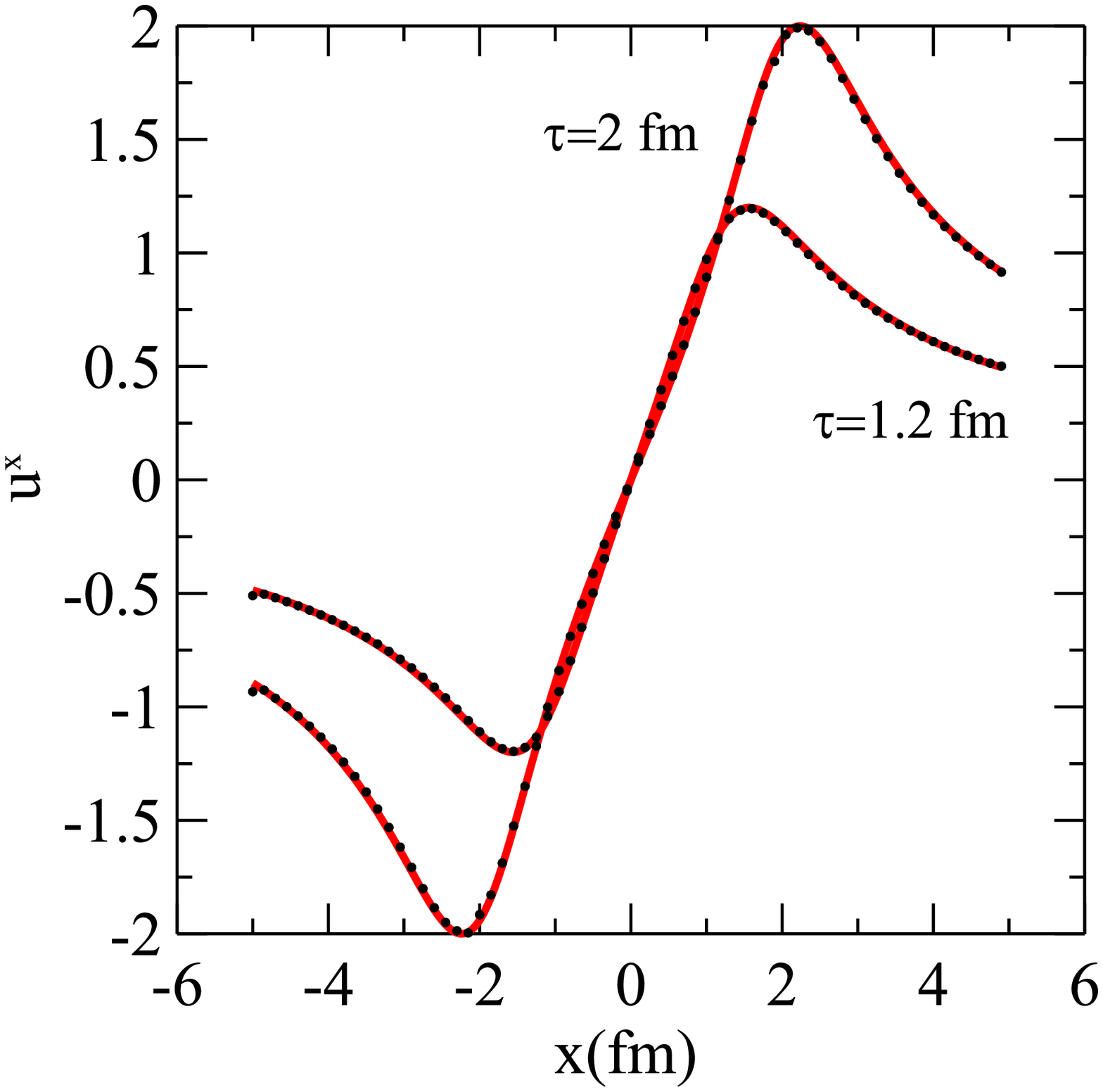} 
\end{minipage} 
\caption{Comparison between the solutions for temperature (left panel) and
the $x$--component of the 4-velocity (right panel) from Gubser flow and 
\textsc{music} (numerical), as a function of $x$. In this plot $\protect\eta%
/s=0.2$ and $\protect\tau_R T=5 \protect\eta/s$. The solid lines denote the
semi-analytic solution while the points denote solutions obtained from 
\textsc{music}.}
\label{MUSIC_T_vel}
\end{figure}

\begin{figure}[tbp]
\begin{minipage}{.4\linewidth} 
\hspace{-1.5cm}
\includegraphics[width=8.2cm]{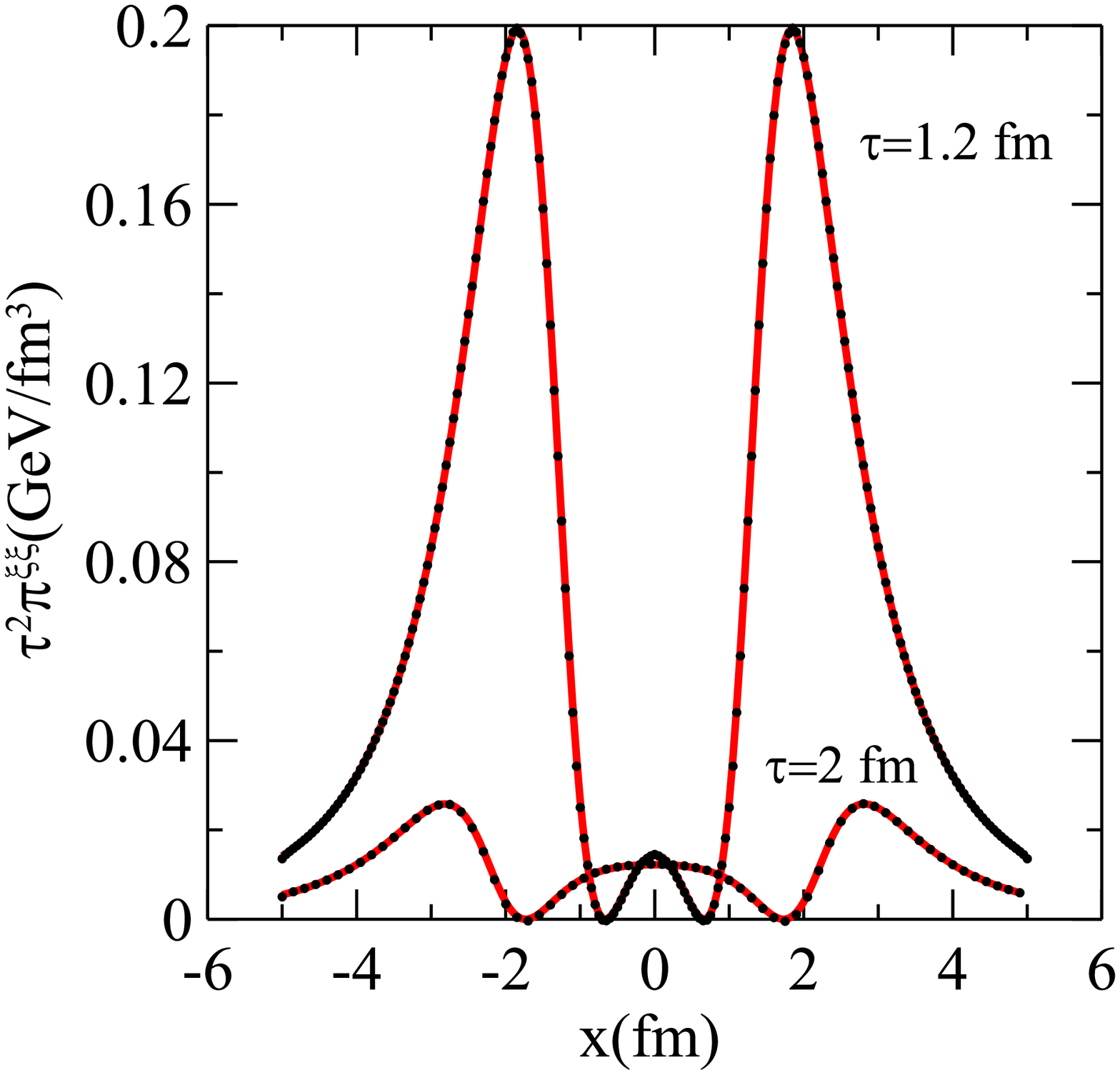} 
\end{minipage}  
\begin{minipage}{.4\linewidth}
\hspace{-1.5cm}
\includegraphics[width=8.2cm]{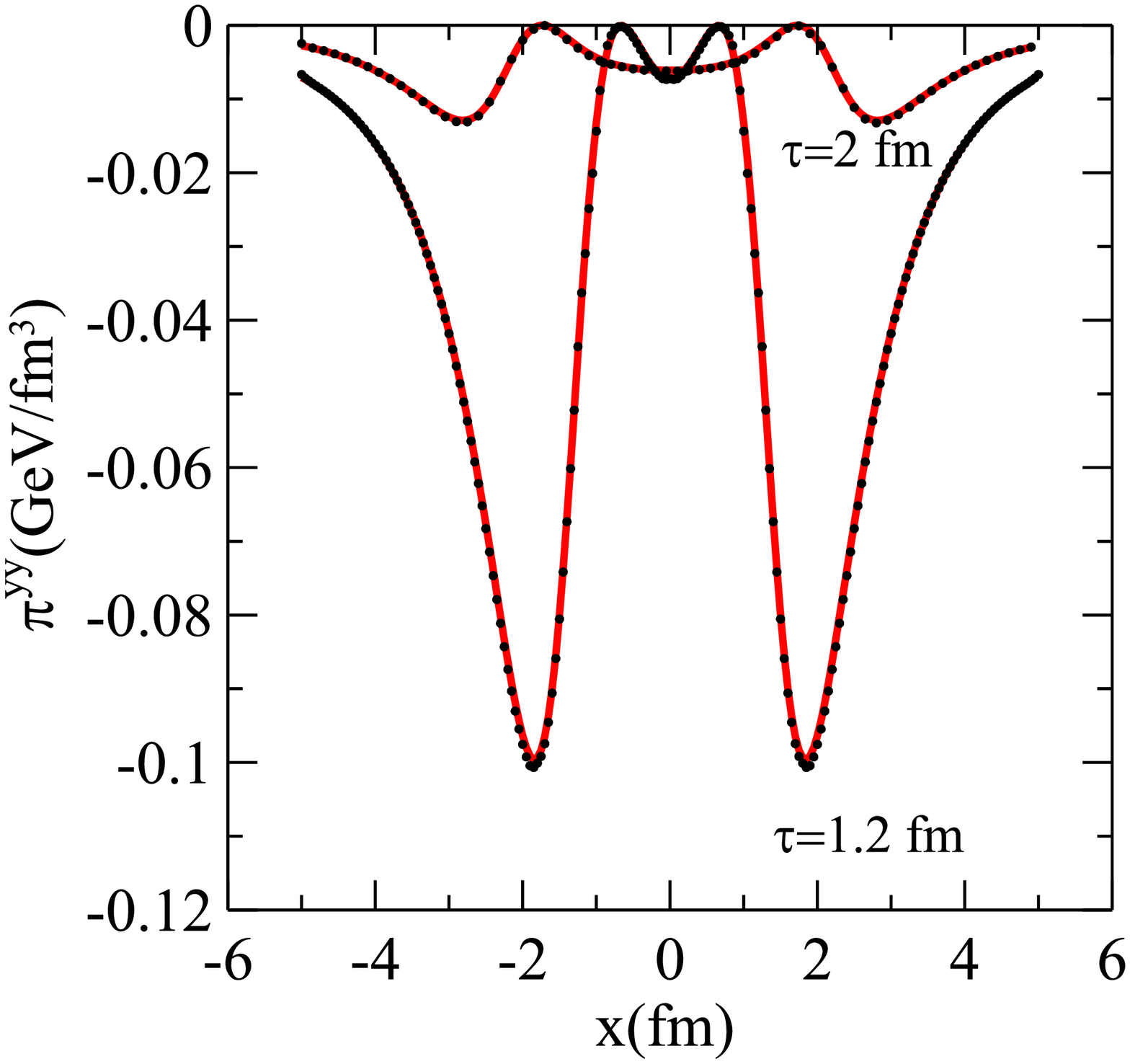} 
\end{minipage}
\begin{minipage}{.45\linewidth}
\hspace{-1.7cm}
\includegraphics[width=8.2cm]{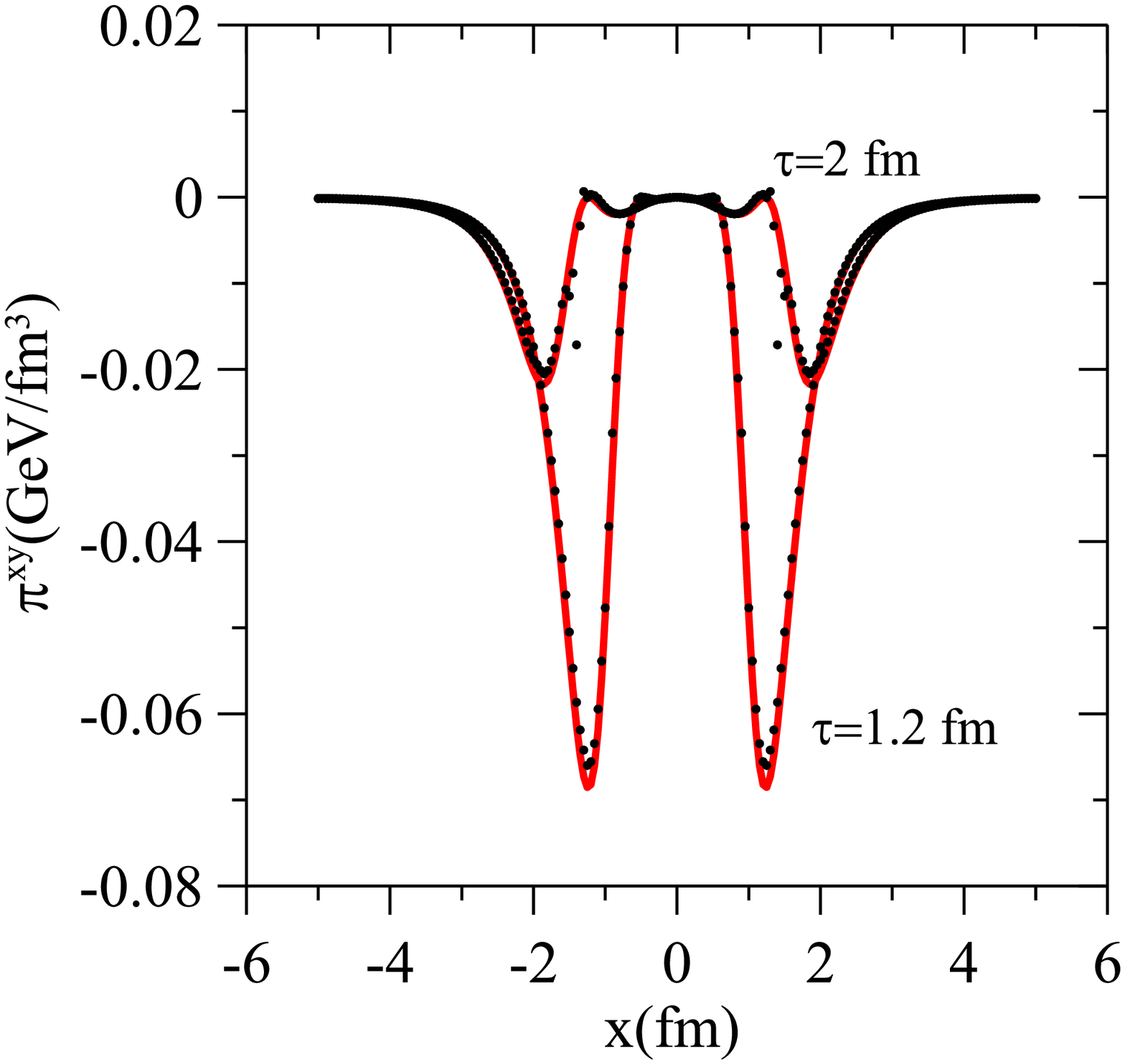} 
\end{minipage}  
\caption{Comparison between the solutions for the $\protect\xi$$\protect\xi$
(left panel), $yy$ (right panel), and $xy$ (lower panel) components of the
shear-stress tensor from Gubser flow and \textsc{music} (numerical), as a
function of $x$. In this plot $\protect\eta/s=0.2$ and $\protect\tau_R T=5 
\protect\eta/s$. The solid lines denote the semi-analytic solution while the
points denote solutions obtained from \textsc{\ music}. }
\label{MUSIC_W}
\end{figure}

One can see that the agreement between the numerical simulation and the
semi-analytical solutions is very good. Only the $xy$ component of the
shear-stress tensor displayed some oscillation at late times. However, since
this component is small, this oscillation is not enough to spoil the overall
agreement.

We remark that such good agreement could only be obtained by adjusting the
flux limiter used in the \textsc{KT} algorithm. Flux limiters are employed
in \textsc{MUSCL} scheme algorithms, such as the \textsc{KT} algorithm, to
control artificial oscillations that usually occur when using higher order
discretization schemes for spatial derivatives. Such spurious oscillations
are known to appear when resolving shock problems, solutions with
discontinuities in density profiles or velocity field, or even when
describing systems which display high gradients, such as the system created
in relativistic heavy ion collisions. Since dissipative effects originate
mainly from space-like gradients of the velocity field, flux limiters are
essential in order to obtain a precise numerical solution of dissipative
fluid dynamics.

\begin{figure}[tbp]
\begin{minipage}{.4\linewidth}
\hspace{-1.5cm}
\includegraphics[width=8.2cm]{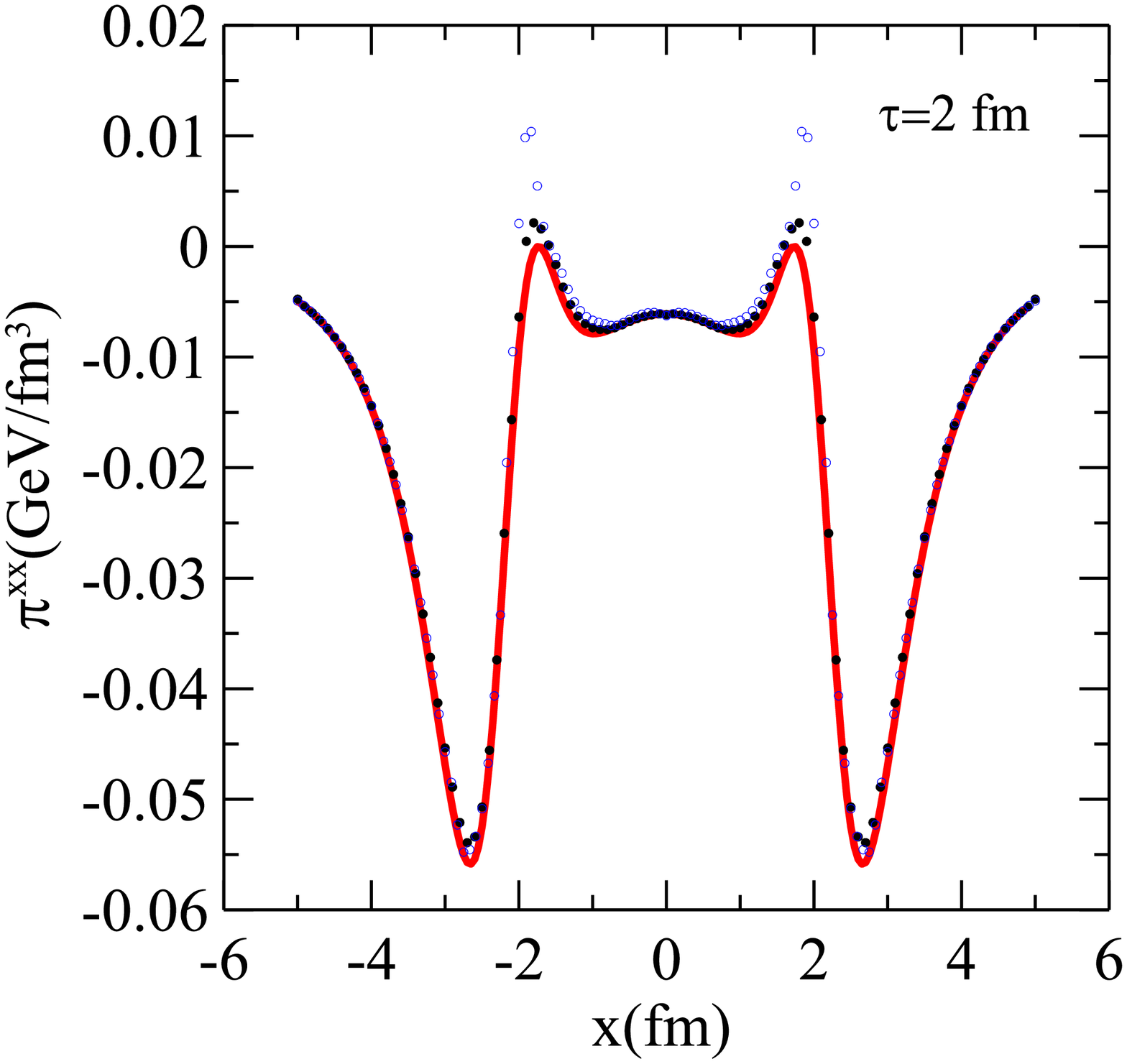} 
\end{minipage}  
\begin{minipage}{.4\linewidth}
\hspace{-1.5cm}
\includegraphics[width=8.2cm]{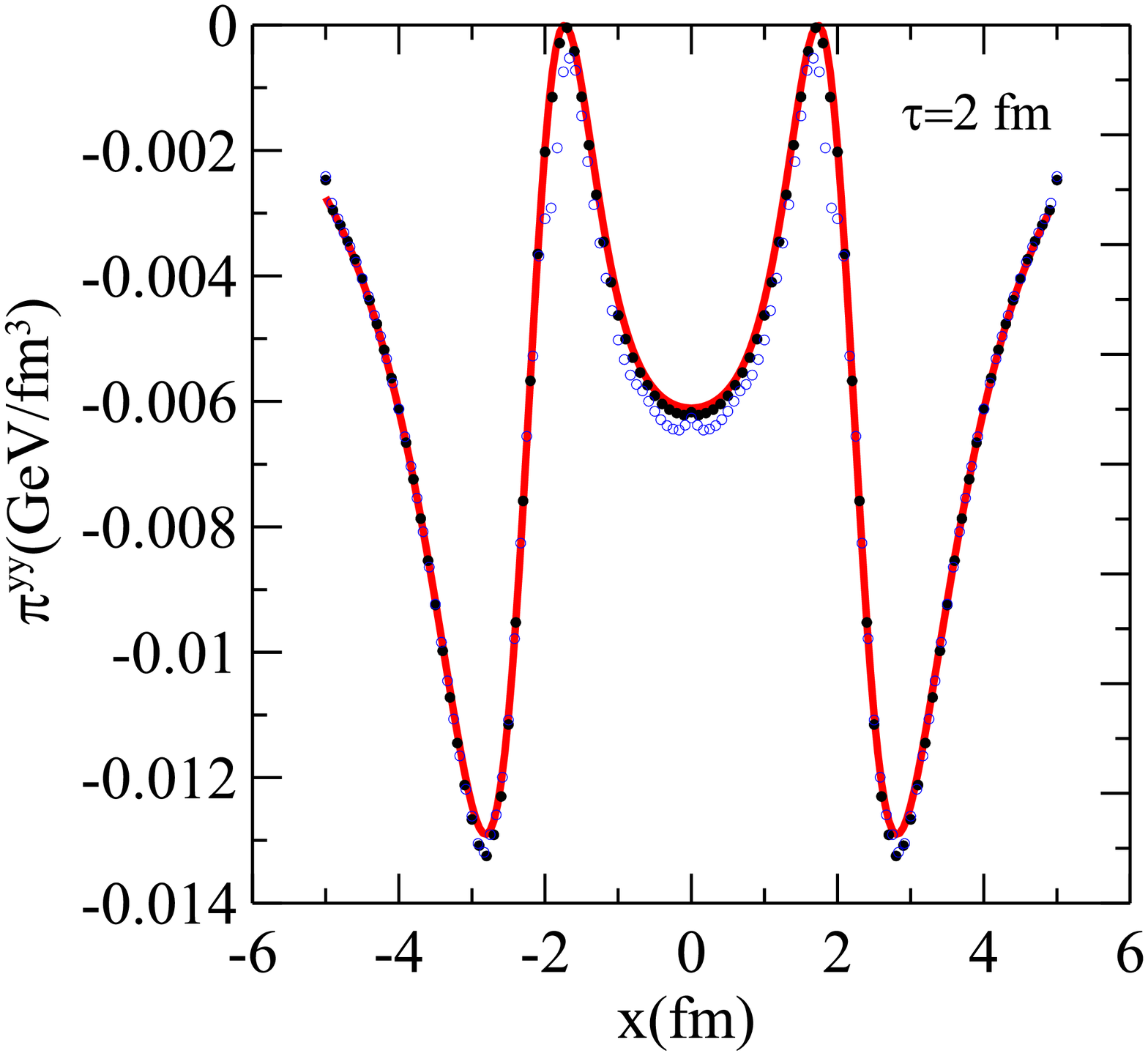} 
\end{minipage} 
\caption{ Numerical solutions of \textsc{music} obtained with $\protect%
\theta =1.1$ (open circles) for the $xx$ (left panel) and $yy$ (right panel)
components of the shear-stress tensor. The full circles correspond to the
solutions obtained with $\protect\theta =1.8$ and the solid lines correspond
to the semi-analytic solution.}
\label{MUSIC_theta}
\end{figure}

Currently, there are several flux limiter algorithms available and many
others still being developed. In \textsc{music}, the van Leer minmod filter
is used~\cite{MUSIC}. In this case, the gradients of currents and fluxes are
controlled according to a free parameter $\theta $, which may vary from $%
\theta =1$ (most dissipative) to $\theta =2$ (least dissipative). The
optimal value of $\theta$ can vary case by case and is usually fixed by
trial and error; in previous work, \textsc{music} was run with $\theta =1.1$%
. However, the agreement displayed in Figs.~\ref{MUSIC_T_vel} and~\ref%
{MUSIC_W} is only obtained by choosing a larger value, $\theta =1.8$,
corresponding to the less diffusive case. The solutions of the temperature
and velocity fields are not very sensitive to changes in the flux limiter
scheme. On the other hand, the solutions of the shear-stress tensor do
depend on the choice of this numerical parameter. In Fig.\ \ref{MUSIC_theta}
we show the numerical solutions of \textsc{music} obtained with $\theta =1.1$
(open circles) for the $xx$ and $yy$ components of the shear-stress tensor,
which are the components most sensitive to this parameter. These solutions
are compared with those of $\theta =1.8$ (full circles) and the
semi-analytical solutions (solid line). One can see that when $\theta =1.1$
the agreement becomes worse, demonstrating the usefulness of the
semi-analytic solution found in this paper in testing the algorithm. It
should be noted that, if a flux limiter is not employed at all, it is not
possible to properly describe the Gubser flow solutions of Israel-Stewart
theory.

\subsection{Comparison to analytical solution}

In the previous section, we showed that an analytical solution for
Israel-Stewart theory can be found in the limit of extremely large viscosity
or, equivalently, of extremely small temperatures (cold plasma limit). Note
that this analytical solution is no longer an approximation if the term $\pi
^{\mu \nu }$ is removed from Israel-Stewart theory. That is, if one solves
the equation,

\begin{equation}
\frac{\tau _{R}}{sT}\left( \Delta _{\alpha }^{\mu }\Delta _{\beta }^{\nu
}\,D_{\tau }\pi ^{\alpha \beta }+\frac{4}{3}\pi ^{\mu \nu }\theta \right) =-%
\frac{2\eta }{s}\frac{\sigma ^{\mu \nu }}{T} ,  \label{SimpleIS}
\end{equation}%
instead of Eq. (\ref{pieq}).

The solution of this equation no longer relaxes to Navies-Stokes theory.
However, it can still be used to test algorithms that solve relativistic
fluid dynamics. The same algorithm that solves Israel--Stewart theory should
also be able to solve the above equation of motion and this can be used as
an independent and powerful test of a given numerical approach. Furthermore,
the term $\pi ^{\mu \nu }$ is rather simple and does not demand much work to
be removed.

\begin{figure}[tbp]
\begin{minipage}{.4\linewidth}
\hspace{-1.5cm}
\includegraphics[width=8.2cm]{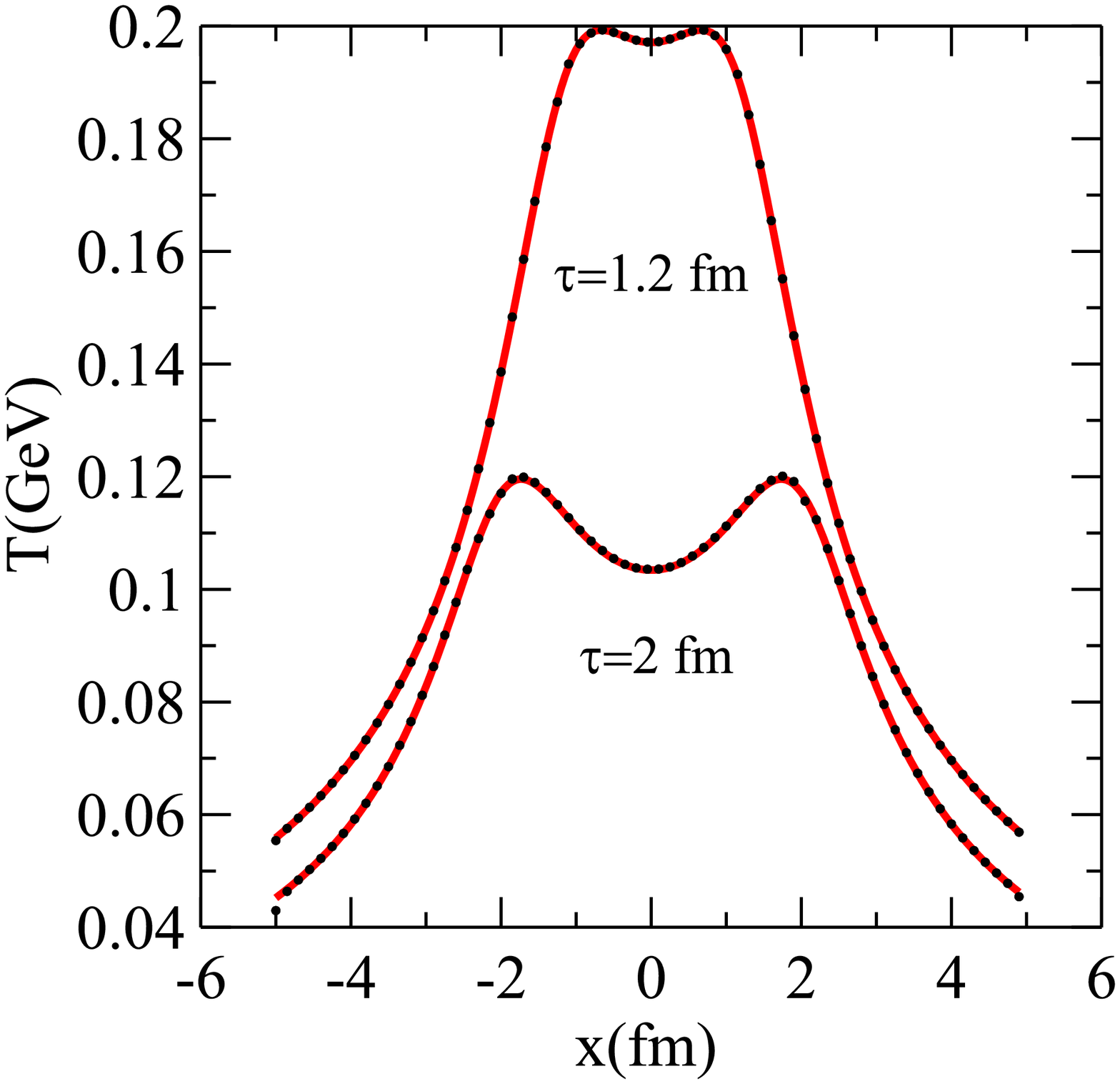} 
\end{minipage}  
\begin{minipage}{.4\linewidth}
\hspace{-1.5cm}
\includegraphics[width=8.2cm]{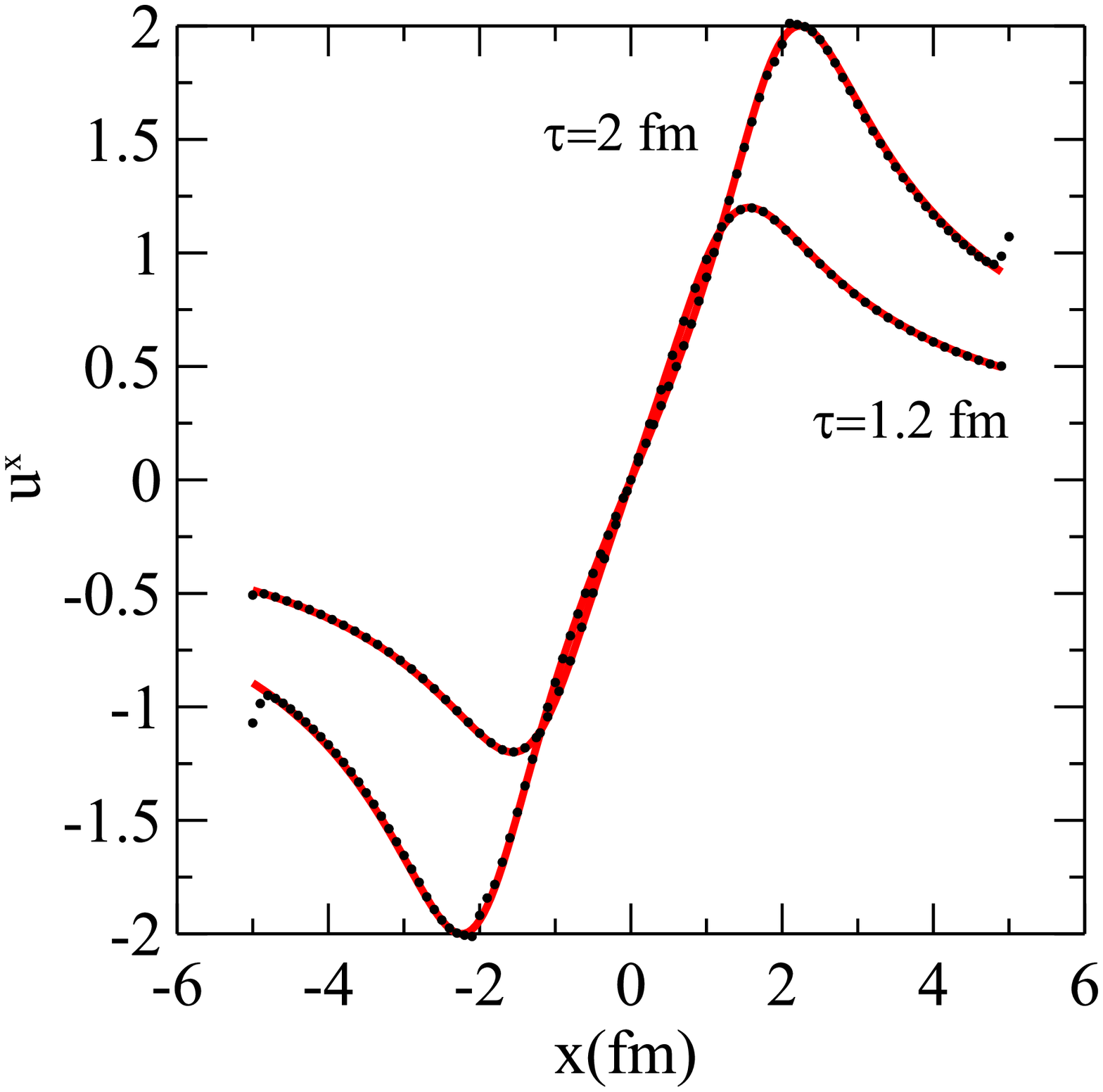} 
\end{minipage} 
\caption{Comparison between the solutions for temperature (left panel) and
the $x$--component of the 4-velocity (right panel) from Gubser flow and 
\textsc{music} (numerical), as a function of $x$. In this plot $\protect\eta%
/s=0.2$ and $\protect\tau_R T=5 \protect\eta/s$. The solid lines denote the
analytic solution while the points denote solutions obtained from \textsc{%
music}.}
\label{MUSIC_T_vel_2}
\end{figure}

\begin{figure}[tbp]
\begin{minipage}{.4\linewidth} 
\hspace{-1.5cm}
\includegraphics[width=8.2cm]{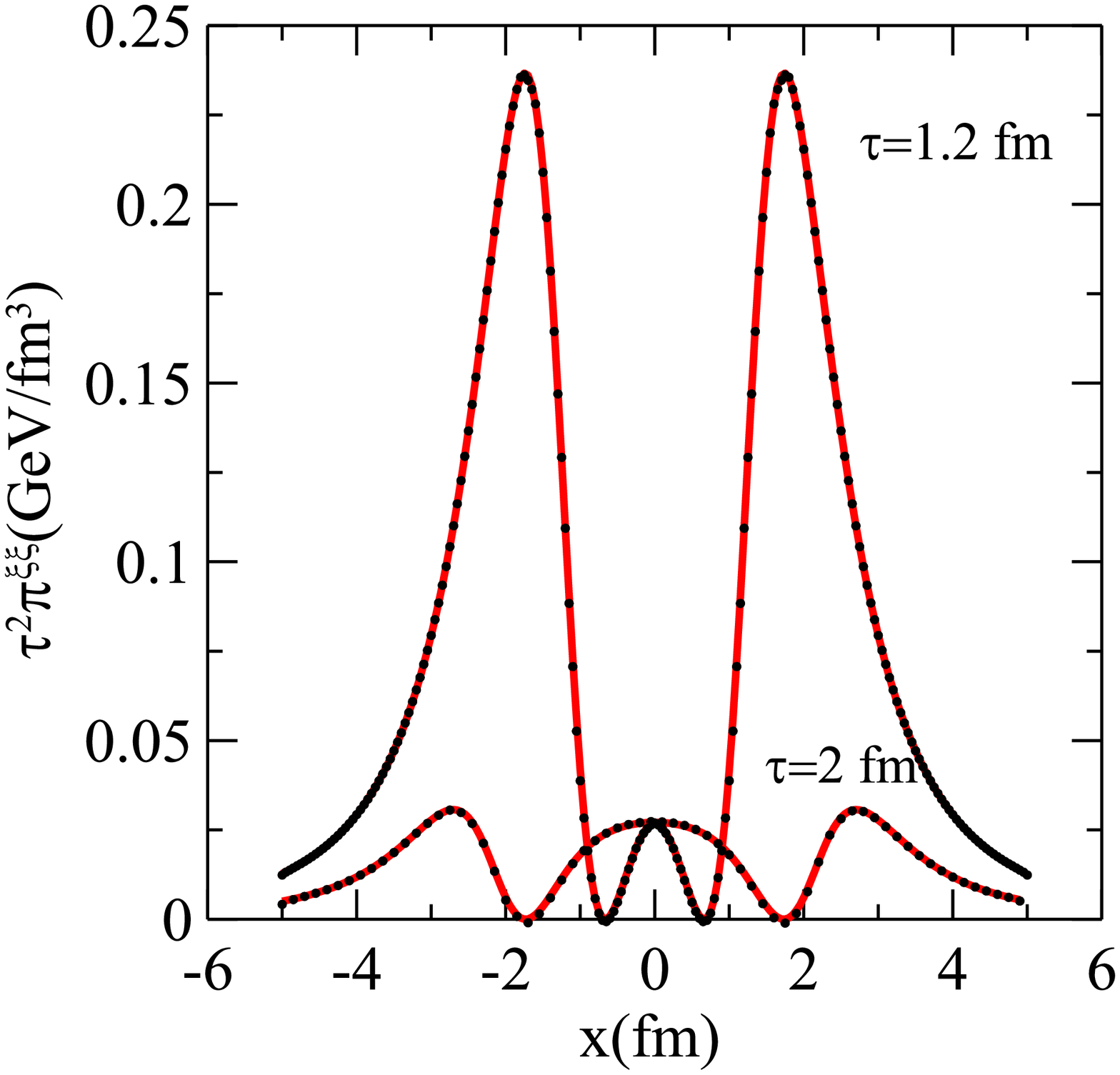} 
\end{minipage}  
\begin{minipage}{.4\linewidth}
\hspace{-1.5cm}
\includegraphics[width=8.2cm]{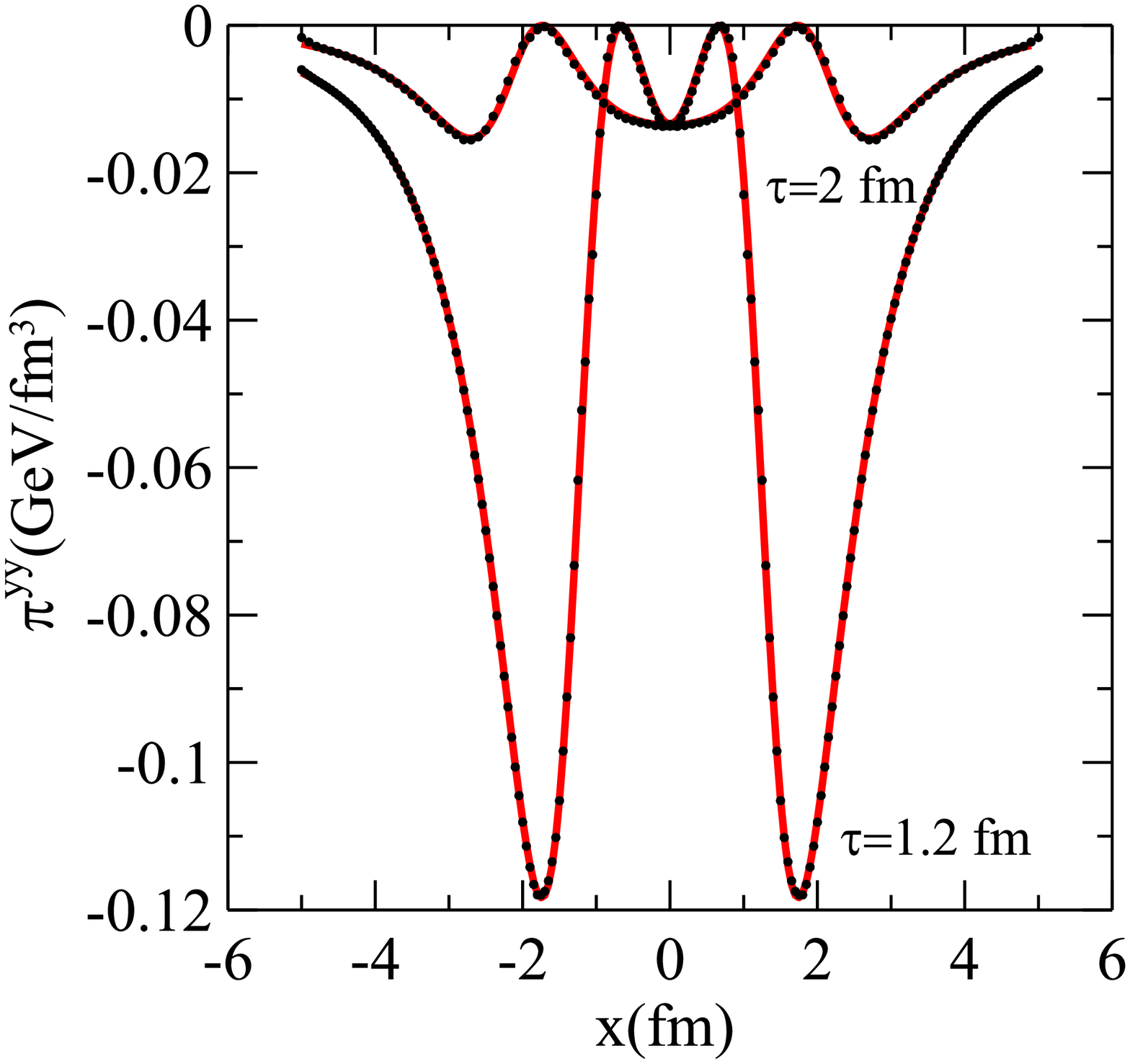} 
\end{minipage}
\begin{minipage}{.5\linewidth}
\hspace{-1.7cm}
\includegraphics[width=8.2cm]{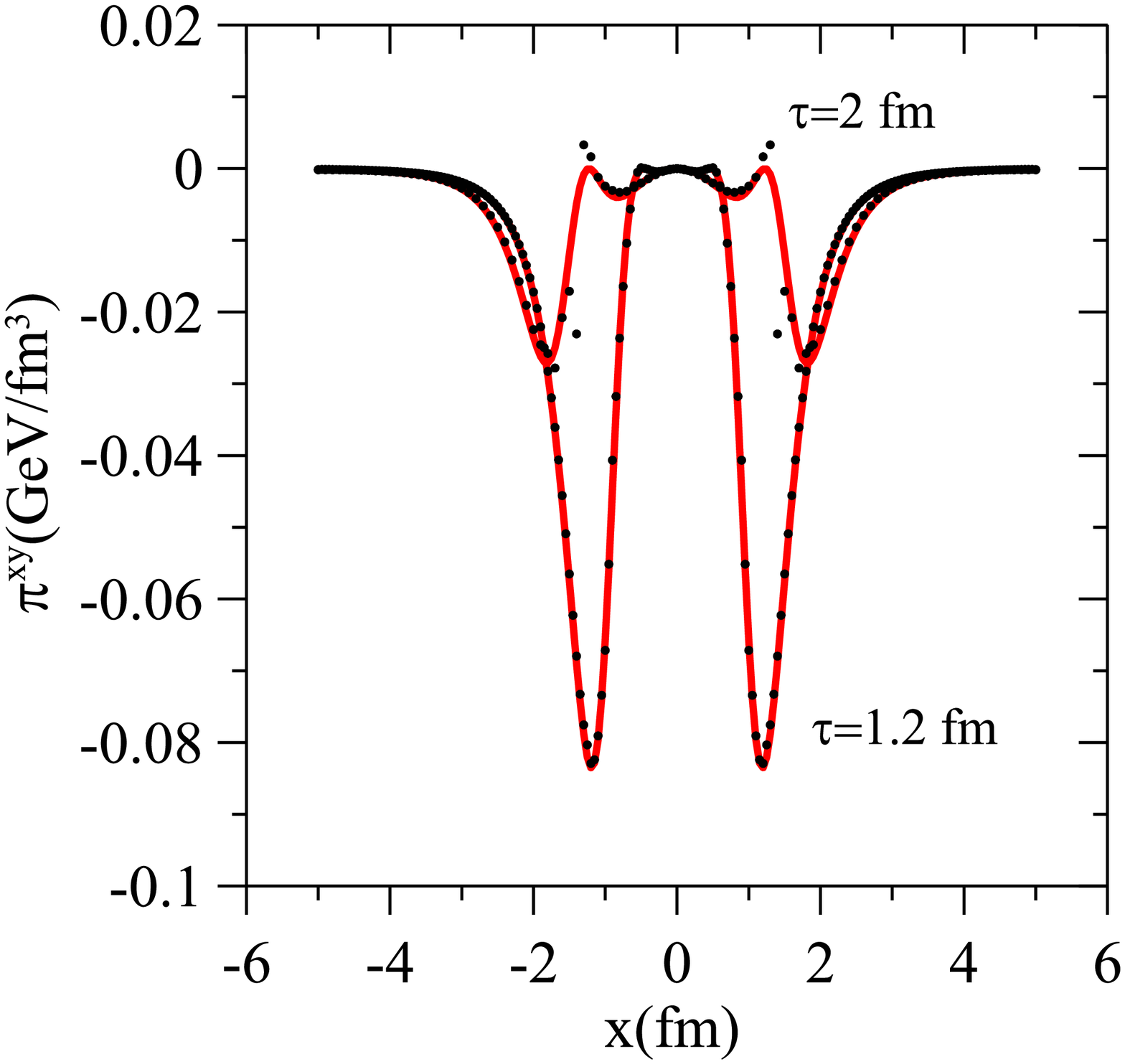} 
\end{minipage}  
\caption{Comparison between the solutions for the $\protect\xi\protect\xi$
(left panel), $yy$ (right panel), and $xy$ (lower panel) components of the
shear-stress tensor from Gubser flow and \textsc{music} (numerical), as a
function of $x$. In this plot $\protect\eta/s=0.2$ and $\protect\tau_R T=5 
\protect\eta/s$. The solid lines denote the analytic solution while the
points denote solutions obtained from \textsc{\ music}. }
\label{MUSIC_W_2}
\end{figure}

As already mentioned, in this case the solution of the theory in de Sitter
space can be found analytically, see Eqs.\ (\ref{analf}) and (\ref{analh}).
We numerically solved Eqs.\ (\ref{energyeq}), (\ref{eulerrel}), and (\ref%
{SimpleIS}) using \textsc{music} by subtracting the aforementioned term,
using the same initial condition described before. The comparison is showed
in Figs.\ \ref{MUSIC_T_vel_2} and \ref{MUSIC_W_2}, which show the spatial
profiles of $T $, $u^{x}$, $\pi ^{\xi \xi }$, $\pi ^{yy}$, and $\pi ^{xy}$.
The solid lines correspond to the analytical solutions while the points
correspond to the numerical solutions of Eq.\ (\ref{SimpleIS}) obtained with 
\textsc{music}.

Note that the level of agreement is the same as before. The solutions in
hyperbolic coordinate even appear to be qualitatively the same, containing
the same general structures as the full solutions. However, from a practical
point of view, the above solutions are very convenient to test a code since
they are already cast in the form of functions and can be written directly
into the code.

\section{Conclusions}

\label{SecC}

We have presented the first analytical and semi-analytical solutions of a
radially expanding viscous conformal fluid that follows relaxation-type
equations such as the Israel-Stewart equations. The $\mathrm{{SO(3)}\otimes {%
SU(1,1)}\otimes {Z}_{2}}$ invariant solutions for the temperature, shear
stress tensor, and flow discussed here can be used to test the existing
numerical algorithms used to solve the equations of motion of viscous
relativistic fluid dynamics in ultrarelativistic heavy ion collision
applications.

We further demonstrated how the solutions derived in this paper can be used
to optimize the numerical algorithm of a well known hydrodynamical code,
fixing numerical parameters that can only be determined by trial and error.
The \textsc{music} simulation code was shown to produce results that are in
good agreement with the analytic and semi-analytic solutions of
Israel-Stewart theory undergoing Gubser flow.

Also, once the temperature and shear-stress tensor profiles are known, one
can use this information for instance to study the energy loss of hard
probes in a radially expanding and viscous QGP scenario \cite{Betz:2013caa,
MARTINI}. Another interesting aspect that could be studied would be the
propagation of small disturbances \cite{gubseryarom,Staig:2011wj} on the
expanding IS fluid background found here in which the temperature is
positive definite throughout the whole dynamical evolution (which is not the
case in the Navier-Stokes solution). Moreover, it would be interesting to see if the
solutions found here for the conformal Israel-Stewart equations correspond to a black
hole configuration in an asymptotically AdS$_{5}$ geometry, as it is the
case for the NS equations at zero chemical potential \cite%
{Bhattacharyya:2008jc}.

The authors acknowledge many useful exchanges with B.~Schenke, and are
grateful to I.~Kozlov, J.-F.~Paquet, D.~H.~Rischke, J.-B.~Rose, and
G.~Vujanovic for discussions. This work was funded in part by Funda\c c\~ao
de Amparo \`a Pesquisa do Estado de S\~ao Paulo (FAPESP), in part by
Conselho Nacional de Desenvolvimento Cient\'ifico e Tecnol\'ogico (CNPq),
and in part by the Natural Sciences and Engineering Research Council of
Canada. G.~S.~Denicol acknowledges the support of a Banting fellowship
provided by the Natural Sciences and Engineering Research Council of Canada.


\end{document}